\def\VUPERarxiv{1}

\documentclass[sigconf]{acmart}

\newif\ifarxiv
\ifdefined\VUPERarxiv \arxivtrue \else \arxivfalse \fi

\copyrightyear{2026}
\acmYear{2026}
\setcopyright{cc}
\setcctype{by}
\acmConference[CCS '26] {Proceedings of the 2026 ACM SIGSAC Conference on Computer and Communications Security}{November 15--19, 2026}{The Hague, Netherlands.}
\acmBooktitle{Proceedings of the 2026 ACM SIGSAC Conference on Computer and Communications Security (CCS '26), November 15--19, 2026, The Hague, Netherlands}
\acmISBN{979-8-4007-2871-6/2026/11}
\acmDOI{10.1145/3830454.3832737}

\usepackage{amsthm}
\usepackage{amsfonts}
\usepackage{listings}

\usepackage{mathtools}
\usepackage{bm}
\usepackage{enumitem}
\usepackage{pifont}
\usepackage{multirow}
\usepackage{subcaption}

\usepackage{soul}

\usepackage{cleveref}

\crefformat{section}{\S#2#1#3}
\crefformat{subsection}{\S#2#1#3}
\crefformat{subsubsection}{\S#2#1#3}

\newif\ifshowcomments
\showcommentstrue

\ifshowcomments
  \newcommand{\syed}[1]{\textcolor{purple}{Syed: #1}}
  \newcommand{\ali}[1]{\textcolor{teal}{[Ali: #1]}}
  \newcommand{\kai}[1]{\textcolor{green}{[Kai: #1]}}
  \newcommand{\gtan}[1]{\textcolor{blue}{GT: #1}}
  \newcommand{\yilu}[1]{\textcolor{brown}{Yilu: #1}}
  \newcommand{\unsure}[1]{\textcolor{red}{unsure: #1}}
  \newcommand{\xiaotian}[1]{\textcolor{red}{Xiaotian: #1}}
\else
  \newcommand{\syed}[1]{}
  \newcommand{\ali}[1]{}
  \newcommand{\kai}[1]{}
  \newcommand{\gtan}[1]{}
  \newcommand{\yilu}[1]{}
  \newcommand{\unsure}[1]{}
  \newcommand{\xiaotian}[1]{}
\fi

\newcommand{\system}{\textsf{VUPER}}

\newcommand{\B}{\mathbb{B}}
\newcommand{\Pos}{\mathbb{P}}

\definecolor{dkgreen}{rgb}{0,.5,0}
\definecolor{ltblue}{rgb}{0,0.4,0.4}
\definecolor{dkviolet}{rgb}{0.3,0,0.5}
\definecolor{dkblue}{rgb}{0,.8,.8}

\lstdefinestyle{asn1style}{
basicstyle=\ttfamily\scriptsize\linespread{1},
mathescape=true,
columns=fullflexible,
xleftmargin=0pt,
xrightmargin=0pt,
keywordstyle=\color{dkblue},
morekeywords={SEQUENCE, CHOICE, SIZE, OF, ENUMERATED},
comment=[l]{--},
morecomment=[l]{--},
commentstyle=\itshape\color{gray},
columns=fullflexible,
keepspaces=true,
moredelim=**[s][\color{dkgreen}]{[[}{]]},
}

\lstdefinestyle{mathstyle}{
basicstyle=\sffamily\selectfont\small,
mathescape=true,
columns=fullflexible,
xleftmargin=0pt,
xrightmargin=0pt,
    morekeywords=[1]{Section, Module, End, Require, Import, Export,
        Variable, Variables, Parameter, Parameters, Axiom, Hypothesis,
        Hypotheses, Notation, Local, Tactic, Reserved, Scope, Open, Close,
        Bind, Delimit, Definition, Let, Ltac, Fixpoint, CoFixpoint, Add,
        Morphism, Relation, Implicit, Arguments, Unset, Contextual,
        Strict, Prenex, Implicits, Inductive, CoInductive, Record,
        Structure, Canonical, Coercion, Context, Class, Global, Instance,
        Program, Infix, Theorem, Lemma, Corollary, Proposition, Fact,
        Remark, Example, Proof, Goal, Save, Qed, Defined, Hint, Resolve,
        Rewrite, View, Search, Show, Print, Printing, All, Eval, Check,
        Projections, inside, outside, Def},
    morekeywords=[2]{forall, exists, exists2, fun, fix, cofix, struct,
        match, with, end, as, in, return, let, if, is, then, else, for, of,
        nosimpl, when},
    morekeywords=[3]{Type, Prop, Set},
    morekeywords=[4]{pose, set, move, case, elim, apply, clear, hnf,
        intro, intros, generalize, rename, pattern, after, destruct,
        induction, using, refine, inversion, injection, rewrite, congr,
        unlock, compute, ring, field, fourier, replace, fold, unfold,
        change, cutrewrite, simpl, have, suff, wlog, suffices, without,
        loss, nat_norm, assert, cut, trivial, revert, bool_congr, nat_congr,
        symmetry, transitivity, auto, split, left, right, autorewrite},
    morekeywords=[5]{by, done, exact, reflexivity, tauto, romega, omega,
        assumption, solve, contradiction, discriminate},
    morekeywords=[6]{do, last, first, try, idtac, repeat},
    morecomment=[s]{(*}{*)},
    identifierstyle={\color{black}},
    keywordstyle=[1]{\color{dkviolet}},
    keywordstyle=[2]{\color{dkgreen}},
    keywordstyle=[3]{\color{ltblue}},
    keywordstyle=[4]{\color{dkblue}},
    keywordstyle=[5]{\color{dkred}},
    commentstyle={\color{dkgreen}},
    aboveskip=3pt, belowskip=3pt
}

\newcommand{\yes}{\ding{108}}
\newcommand{\no}{\ding{109}}
\newcommand{\limited}{\textcolor{gray}{\ding{108}}}

\newcommand{\diagcell}[2]{%
  \raisebox{-4.5pt}{%
    \rlap{\setlength{\unitlength}{1pt}%
          \raisebox{-2.2pt}{\begin{picture}(50,17)(0,0)
            \qbezier(0,17)(25,8.5)(50,0)
          \end{picture}}}%
    \shortstack{\makebox[50pt][r]{#2}\\[1pt] \makebox[50pt][l]{#1}}%
  }%
}

\ifarxiv
  \newcommand{\extref}[1]{Appendix~\ref{#1}}
  \newcommand{\Extref}[1]{Appendix~\ref{#1}}
\else
  \newcommand{\extref}[1]{the extended version~\cite{vuper-extended}}
  \newcommand{\Extref}[1]{The extended version~\cite{vuper-extended}}
\fi

\newcommand{\impact}{\noindent\textbf{\textit{Impact.}} }

\newcounter{ccount}

\newcommand{\cparagraph}[1]{%
  \stepcounter{ccount}
  \phantomsection
  \noindent\textbf{C\theccount: #1.}\label{challenge:\theccount}
}

\newcommand{\para}[1]{%
  \phantomsection
  \noindent\textbf{#1.}
}

\newcommand{\smallpara}[1]{
\noindent{\textit{\underline{#1.}}}
}

\newcommand{\challengeref}[1]{\hyperref[challenge:#1]{\textbf{C#1}}}


\setlist{topsep=0pt, partopsep=0pt, itemsep=0pt, parsep=0pt}
\setlist[description]{topsep=0pt, partopsep=0pt, itemsep=0pt, parsep=0pt, leftmargin=.75em}

\begin{document}

\title{\system{}: Verified ASN.1 UPER Parser}

\author{Xiaotian Zhou}
\email{xzz5503@psu.edu}
\affiliation{%
  \institution{The Pennsylvania State University}
    \city{University Park}
  \state{PA}
  \country{United States}
}

\author{Kai Tu}
\email{kjt5562@psu.edu}
\affiliation{%
  \institution{The Pennsylvania State University}
  \city{University Park}
  \state{PA}
  \country{United States}
}

\author{Ali Ranjbar}
\email{aranjbar@psu.edu}
\affiliation{%
  \institution{The Pennsylvania State University}
  \city{University Park}
  \state{PA}
  \country{United States}
}

\author{Yilu Dong}
\email{yiludong@psu.edu}
\affiliation{%
  \institution{The Pennsylvania State University}
  \city{University Park}
  \state{PA}
  \country{United States}
}

\author{Gang Tan}
\email{gtan@psu.edu}
\affiliation{%
  \institution{The Pennsylvania State University}
  \city{University Park}
  \state{PA}
  \country{United States}
}

\author{Syed Rafiul Hussain}
\email{hussain1@psu.edu}
\affiliation{%
  \institution{The Pennsylvania State University}
  \city{University Park}
  \state{PA}
  \country{United States}
}

\renewcommand{\shortauthors}{Xiaotian Zhou et al.}

\begin{abstract}
ASN.1 is a widely used interface description language, and UPER (Unaligned Packed Encoding Rules) is one of its key encoding rules, particularly popular in security-critical domains such as cellular networks and vehicle-to-everything (V2X) communication. To ensure the correctness and security of this foundational infrastructure, we present VUPER, a framework for generating verified ASN.1 UPER parsers. We first formalize the notion of a bit-precise parser and identify properties that prove round-trip consistency for parsers and serializers, while accounting for ASN.1 features such as backward/forward compatibility. We then implement and verify parser and serializer combinators for ASN.1 basic types and structures, while adhering to the UPER specification. We also develop a compiler that translates ASN.1 definitions into verified parsers. Finally, we develop a dynamic testing framework using the VUPER parser as a test oracle. 
To empirically evaluate our approach, we test 7 open-source and 4 commercial ASN.1 parsers using 5G and V2X communication protocols. VUPER uncovers 20 types of inconsistencies in popular parsers and demonstrates stricter compliance with ASN.1 UPER standards. Additionally, we demonstrate concrete attacks by exploiting these parser vulnerabilities.  
\end{abstract}



\maketitle

\section{Introduction} \label{intro}

Parsing and serializing binary data formats are fundamental tasks in network protocols. The parser must ensure the correct reconstruction of the original message encoded by the serializer and safely reject potentially malicious inputs.
Interface description languages (IDL) streamline this process by
decoupling parsing from the main program, thus allowing the reuse of parser libraries across protocols sharing the same IDL. 
Abstract Syntax Notation One (ASN.1) is a widely used IDL standardized by the International Telecommunication Union (ITU)~\cite{ITU_X680}. 
Among ASN.1 encoding rules, such as Basic Encoding Rules or BER, Distinguished Encoding Rules or DER, Unaligned Packed Encoding Rules (UPER) are most widely used in security-critical wireless protocols such as cellular networks and vehicle-to-everything (V2X) systems due to their efficiency, compactness, expressiveness, and interoperability. Despite their prevalence and complexity, UPER implementations remain under-investigated \cite{tullsen2019asn} and lead to critical vulnerabilities, ranging from safety violations to subtle logic bugs \cite{cve2019-6740}. 

One of the main reasons for such flaws is that the ASN.1 standard exists only as descriptive text, lacking a precise reference implementation. However, the intricate nature of ASN.1, including its support for various primitive types and combinators, ranges, optional fields, extensions, and the bit-level precision required by UPER, makes it very difficult to derive a fully correct implementation. This leaves the ASN.1 implementations highly error-prone,
as evidenced by the over 100 reported CVEs~\cite{cve_asn1}.
Subtle flaws or non-conformance in the parser can also propagate issues throughout the downstream systems, potentially leading to authentication bypass~\cite{cve_x509_auth_bypass}, sensitive information exposure~\cite{cve_2024_5535_leak_info}, or denial-of-service (DoS)~\cite{ransack24} attacks.

Prior work on securing parsers has primarily focused on testing and fuzzing \cite{hernandez_firmwire_2022}. Although these approaches are effective for identifying vulnerabilities, they cannot provide the formal guarantees that eliminate entire classes of errors. In contrast, constructing parsers with provable security guarantees can achieve this by design~\cite{Everparse19, wallez2023comparse}.
Such guarantees are typically based on a \textit{round-trip} property~\cite{Everparse19, wallez2023comparse}, which requires the parser and serializer to be mathematical inverses. This property ensures two critical principles: \textit{non-ambiguity}, where a given encoding maps to at most one message, and \textit{non-malleability}, where a message has only one canonical encoding. Enforcing these principles ensures that the parser correctly interprets the input while preventing all malformed ones.

Previous efforts on verifying binary formats cover both generic formats~\cite{Narcissus19, Everparse19, geest17generic}, and specific formats such as ASN.1 DER~\cite{ni2023asn1}, and  Protocol Buffer~\cite{ye19verified}. 
However, when it comes to ASN.1 UPER, prior approaches have several critical limitations (summarized in Table \ref{tab:verified_parser_comparsion}).
First, specifying the correctness property for UPER is complicated by the protocol's flexibility. For instance, EverParse~\cite{Everparse19} framework enforces a strict bijective mapping between parser and serializer, but this property cannot easily accommodate the extensibility requirements of UPER. 
Meanwhile, ignoring malleability \cite{geest17generic, asn1acn25bucev} entirely is overly permissive and risks overlooking malformed packets. 
Second, most frameworks are byte-oriented~\cite{Everparse19, wallez2023comparse, vest_25}, creating a mismatch when applied to the bit-precise protocol UPER. 
Furthermore, high-performance zero-copy systems like EverParse~\cite{Everparse19} cannot be easily ported to UPER, because bit-precise protocols mandate explicit value extraction from arbitrary bit offsets. 
Adapting these architectures would require a complete redesign and likely negate their original performance advantages.
Finally, most existing ASN.1 formalizations focus on DER~\cite{ni2023asn1, asn1acn25bucev, debnath2024armor, verdict_x509_25}, whereas the only prior work on UPER~\cite{tullsen2018cav} lacks support for the critical yet most common UPER features, such as \textsf{CHOICE} types and \textsf{extensions} while imposing a hard-coded limit (e.g., 37 bytes) on input.

To bridge these gaps, we present \system{}, the \emph{first comprehensive} framework for representing bit-precise parsers and serializers and formally verifying the correctness of ASN.1 UPER parsers, using the Rocq prover~\cite{Coq-refman}. \system{} achieves data compactness by using a bitstream model that disregards byte alignment, while still maintaining reasonable performance. 
\system{} guarantees a nuanced round-trip property between its parser and serializer to account for forward and backward compatibility. For messages without unknown extensions, a strict injection property ensures full non-malleability. Conversely, for messages with unknown extensions, the weak-injection property applies a more relaxed criterion: the parsed message can be successfully re-encoded, without insisting that the output exactly match the input.
Together, these properties demonstrate that the parser rejects malformed or malicious inputs and remains correct with respect to the serializer.
We also prove auxiliary properties essential for building other guarantees, including \textit{local write} and \textit{encode-consistency}. 
We adopt a \textit{modular} parser-combinator approach. We develop a family of parsers and serializers that range from bit-level manipulation to structural types like \textsf{CHOICE} and \textsf{SEQUENCE}. 
Each component is formally verified to ensure round-trip consistency. By utilizing these combinators, we build a compiler that automatically transforms ASN.1 definitions into verified parser/serializer pairs. 
 
We further develop a testing framework using our formally verified \system{} parser (extracted from Rocq to OCaml) as a trusted oracle.
The framework helps systematically uncover discrepancies in other ASN.1 implementations. 
To effectively generate testing inputs, we use grammar-aware mutation by incorporating ASNFuzzGen~\cite{ransack24} into the fuzzing loop. To ensure precise comparison, we utilize the JSON Encoding Rules (JER)~\cite{ITU_X697} to normalize and compare the values produced by each parser.

We evaluated 11 parser libraries by testing them with uplink and downlink messages from 5G and Intelligent Transport Systems (ITS) protocols while using the verified parser as the oracle. Our evaluation includes 7 open-source and 4 closed-source implementations, notably two ASN.1 UPER parsers from commercial 4G and 5G basebands.
We find that \system{} enforces stricter compliance than most existing implementations and uncovers 20 types of inconsistencies in other libraries, including lenient constraint checks, value corruption, and deviations from the UPER standard. 
To further evaluate the security implications of these defects, we demonstrate 4 concrete attacks against two widely used open-source 4G/5G stacks. These exploits include DoS via unfiltered inputs and cryptographic key misconfigurations due to parser misinterpretation.
We also benchmark the performance of the {\system}-generated parser on 5G messages, which is on average 4 times slower than C libraries~\cite{ASN1c, titan_asn1}, and 4 times faster than Python libraries  ~\cite{pycrate}.
This is expected as \system{} prioritizes formal correctness over efficiency and  
is intended to be used as a reference implementation for testing other ASN.1 parsers and fixing bugs. 
This is a key first step toward a performant, low-overhead, and formally verified ASN.1 UPER parser. 

\para{Contributions} 
We make the following contributions.  
\begin{itemize}[leftmargin=*]
    \item 
        We design \system{}, an automated, bit-precise, and correct-by-construction ASN.1 UPER parser and serializer library.
    
    \item 
        We prove and implement the \textit{round-trip} properties for extensions, and the basic and ASN.1 parser combinators of our parser.

    \item 
        We develop a testing framework based on \system{} as a trusted oracle to effectively validate existing UPER parsers by leveraging \system{}'s formal guarantees. 
    
    \item
        We evaluate 7 open-source and 4 commercial parser implementations and uncovered 20 types of non-compliance, which can lead to sensitive information exposure and DoS. 
\end{itemize}

\section{Background} \label{sec:background}

\noindent
\textbf{Abstract Syntax Notation One} (ASN.1) 
is a formal language for defining the vendor-neutral data structures used in telecommunication protocols~\cite{LTE_RRC, CAM_ITS, ITU_X509}.
Its primary strength lies in decoupling abstract data logic from its physical bit/byte representation, ensuring interoperability across diverse systems. 
ASN.1 provides a highly expressive syntax for this purpose.
Figure~\ref{subfig:asn1_def_b} shows an ASN.1 definition of \textsf{PDSCH-ServingCellConfig} from the 5G Radio Resource Control (RRC) protocol~\cite{5gRRC_NR}. The message is transmitted from 5G base stations to user equipment (UE), and contains parameters necessary for processing the Physical Downlink Shared Channel (PDSCH) on the corresponding cell. 
The message is a \textsf{SEQUENCE} (i.e., struct) of fields and contains primitive types like \textsf{BOOLEAN} and \textsf{INTEGER} and user-defined types like \textsf{ServCellIndex}. It also includes parameterized types
and markers (e.g., \textsf{OPTIONAL}) for fields to describe their behavior.

\begin{figure}[t]
    \centering
    \begin{subfigure}{\linewidth}
        \begin{lstlisting}[style=asn1style,aboveskip=0pt, belowskip=0pt]
PDSCH-ServingCellConfig ::= SEQUENCE {
  codeBlockGrpTx    SetupRelease { PDSCH-CodeBlockGrpTx }  OPTIONAL, 
  xOverhead         ENUMERATED { xOh6, xOh12, xOh18 }      OPTIONAL, 
  nrofHARQ          ENUMERATED {n2, n4, n6, n10, n12, n16} OPTIONAL, 
  pucch-Cell        ServCellIndex                          OPTIONAL, 
  ... }
\end{lstlisting}
    \caption{Definition in RRC Release 15.2.1}
    \label{subfig:asn1_def_a}
    \end{subfigure}

\begin{subfigure}{\linewidth}
    \begin{lstlisting}[style=asn1style,aboveskip=3pt, belowskip=0pt]
PDSCH-ServingCellConfig ::= SEQUENCE {
  codeBlockGrpTx    SetupRelease { PDSCH-CodeBlockGrpTx }  OPTIONAL, 
  xOverhead         ENUMERATED { xOh6, xOh12, xOh18 }      OPTIONAL, 
  nrofHARQ          ENUMERATED {n2, n4, n6, n10, n12, n16} OPTIONAL, 
  pucch-Cell        ServCellIndex OPTIONAL, 
  ...,
  [[ maxMIMO-Layers           INTEGER (1..8)               OPTIONAL, 
     processingType2Enabled   BOOLEAN                      OPTIONAL ]], 
  [[ pdsch-CodeBlockGrpTxList-r16
            SetupRelease { PDSCH-CodeBlockGrpTxList-r16 } OPTIONAL
  ]] } 
\end{lstlisting}
\caption{Definition in RRC Release 16.3.1}
\label{subfig:asn1_def_b}
\end{subfigure}

    \caption{ Definition of \textsf{PDSCH-ServingCellConfig} in two different releases with some field and type names simplified. }
    \label{fig:asn1_example1}
\end{figure}

\begin{figure}
    \centering
    \begin{subfigure}{0.48\linewidth}
        \begin{lstlisting}[style=asn1style,aboveskip=0pt, belowskip=0pt]
tmobile-PDSCH-ServingCellConfig ::= {
  nrofHARQ        n16,
  maxMIMO-Layers  2 
} 
        \end{lstlisting}
        \caption{Message from Newer Version (Rel 16.3.1)}
        \label{subfig:example_msg_a}
    \end{subfigure}
    \hfill
    \begin{subfigure}{0.48\linewidth}
        \begin{lstlisting}[style=asn1style,aboveskip=0pt, belowskip=0pt]
tmobile-PDSCH-ServingCellConfig ::= {     
  nrofHARQ        n16
  -- Skipped unknown field
} 
        \end{lstlisting}
        \caption{Message After Decoded by Older System (Rel 15.2.1)}
        \label{subfig:example_msg_b}
    \end{subfigure}
    \begin{subfigure}{\linewidth}
        \includegraphics[width=\linewidth]{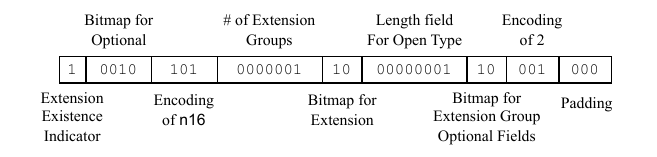}
        \caption{Bit Encoding of Message in (a)}
        \label{subfig:example_msg_big_encoding}
    \end{subfigure}
        \caption{Concrete \textsf{PDSCH-ServingCellConfig} Message (Rel. 16.3.1) sent by T-Mobile Network with its Bit Encoding}
    \label{fig:bit_encoding}
\end{figure}

\para{Extensions} The example in Figure~\ref{fig:asn1_example1} also illustrates ASN.1 extensions. The definition in Figure~\ref{subfig:asn1_def_a} shows the base message from an earlier 5G release, future-proofed with the `\texttt{...}' extension marker. 
The definition in Figure~\ref{subfig:asn1_def_b} is a later release with new capabilities and adds new fields such as \textsf{maxMIMO-Layers}. These additions are organized into ExtensionAdditionGroups, wrapped by the double square brackets ([[...]]).
Extensions are essential for ensuring interoperability, enabling systems operating on different releases to communicate effectively without errors. 
For forward compatibility, if an old system (e.g., Rel-15.2.1) receives a message like Figure~\ref{subfig:example_msg_a} from a newer system (e.g., Rel-16.3.1), it would process known fields such as \textsf{nrofHARQ}, while skipping the unknown fields like \textsf{maxMIMO-Layers}, resulting in the decoded messages in Figure~\ref{subfig:example_msg_b}. 
Similarly, backward compatibility means the newer system should handle the absence of certain fields gracefully.
Furthermore, ASN.1 extensions must not be confused with messages labeled as `extension' or other protocol-specific extension mechanisms. To highlight the essential role of extensions in complex networks, there are more than 500 occurrences of extension markers out of a total of 2,500 defined message types in 5G RRC (Rel 17)~\cite{5gRRC_NR}.

\para{UPER (Unaligned Packed Encoding Rules)}
Once the abstract type is defined, the value for each message can be physically represented using one of several mutually incompatible encoding rules, each suited for different purposes.

\smallpara{Compact Encoding} 
Rules of BER/DER use a byte-aligned tag-length-value (TLV) format. In contrast, Packed Encoding Rules (PER) prioritize compactness, with UPER going furthest by eliminating all byte alignment. 
For example, a constrained type like \textsf{INTEGER(1..8)} is encoded in UPER using only 3 bits instead of 4 bytes (32-bit int) to represent values from 1 to 8. 
Such a setup is perfect for bandwidth-limited environments. For instance, 
the RRC protocols in 4G/5G~\cite{5gRRC_NR} use UPER for efficient over-the-air messaging. UPER is likewise employed in Intelligent Transport Systems (ITS) standards~\cite{SAEJ2735_2020, CAM_ITS}, encoding messages from vehicle-to-vehicle communication to traffic management.

\smallpara{Canonical vs. Basic UPER}
Canonical UPER enforces a mandatory 1-1 correspondence between the value and its bit representation, while Basic is lenient on some occasions.

\smallpara{Illustration of UPER's bit-level encoding} 
Figure~\ref{subfig:example_msg_big_encoding} illustrates the bit-level encoding of a concrete \textsf{PDSCH-ServingCellConfig} message defined in Figure~\ref{subfig:example_msg_a}.
The encoding begins with a one-bit flag to indicate the presence of extensions, followed by a presence bitmap for the optional fields for the \textit{sequence root} (items before the extension marker). In this message, the bitmap is $0010$ because only the 3rd field out of the 4, \textsf{nrofHARQ}, is present. The encoded values of all present fields appear sequentially after the bitmap.
Next is the \textit{extension encoding}, which starts with a length determinant, a binary value indicating the number of extension groups. In this case, the value 0000001 represents 2 extension groups (the count starts at 1 in this encoding scheme). This is followed by a bitmap indicating which extension groups are present. Here, since only the first group exists, the bitmap is 10.
Each group carries its own presence bitmap and field values. 
Here, the first extension group only includes the first field \textsf{maxMIMO-Layers} out of the two, so its bitmap is 10. The value of that field follows.
Finally, the extension group is wrapped in an Open Type field.
During encoding of an Open Type field, the inner type is first generated and padded to the nearest octet boundary. A length determinant is then prefixed to specify the total number of octets, followed by the octet-aligned payload. 
This structure allows a parser to safely skip unrecognized extensions by advancing the bitstream by the indicated octet length, which facilitates forward compatibility.

\section{Design Principles of \system} \label{sec:design_principle}

To present the design principles of \system{}, we first begin with the parser's round-trip  properties, then turn to our verification approaches, and finally discuss the technical challenges.

\subsection{Properties of Parsers} 
\label{sec:high_level_parser_prop}
In general, parser security is often framed in terms of two dual properties: \textit{surjection} and \textit{injection}~\cite{Everparse19, wallez2023comparse, tullsen2018cav}. 
A serializer $s : A \to \B$ maps data of type $A$ into its encoded form $\B$, usually a byte array. And a parser $p : \B \to A \cup \bot$ does the reverse, converting an encoded input back to the value in $A$, or returning $\bot$ for invalid inputs. 

\begin{description}[leftmargin=.75em]
    \item[\textit{Surjection.}]  A parser $p$ is a surjective inverse of a serializer $s$ if, for any message $a \in A$, it satisfies $p(s(a)) = a$. Surjection also ensures non-ambiguity: no two distinct messages share the same encoding. Formally, $\forall a, a' \in A$, if $s(a) = s(a')$, then $a = a'$.

    \item[\textit{Injection.}] It requires that the serializer recovers exactly the original encoding whenever parsing succeeds. Formulated as $\forall b \in \B$, $s(p (b)) = b $ or $p(b) = \bot$. This implies non-malleability: for all $b, b' \in \B$, if $p(b) = p(b')$, then either $b = b'$ or $p(b) = \bot$, ensuring each valid encoding corresponds to a unique message. 
\end{description}
These properties align with Canonical UPER, which enforces a strict bijective mapping of values and bitstrings. 
The \textit{surjection} property ensures that every encoded value can be correctly parsed. 
This property establishes the \textit{completeness} of the parser relative to the serializer, ensuring it never rejects correctly encoded inputs.
The \textit{injection} property can prevent security issues by properly filtering all malformed inputs~\cite{Everparse19, wallez2023comparse}, such as out-of-range values or non-canonical encodings.
\textit{Termination} is also crucial for program reliability by preventing infinite loops that may cause freezes or resource exhaustion.
We exclude side-channel security~\cite{vest_25} as enforcing constant-time execution is generally impractical for ASN.1 UPER, where payloads range from simple values to complex nested structures~\cite{5gRRC_NR}.
We also exclude input exhaustion, as trailing bits do not affect the semantic correctness of a successfully decoded value.

\para{Threat Model}
Since ASN.1 parsers are used in many network protocols, a remote attacker could exploit parser vulnerabilities by transmitting specially-crafted payloads to the victim's network interface. For instance, the RRC protocol~\cite{5gRRC_NR, LTE_RRC} between the UE and gNB exposes 3 possible attack scenarios.
\ding{182} A \textit{fake base station} can lure the UE to connect and send malicious RRC messages to the UE, causing DoS of the user \cite{tu2024logic, shaik2019new, park2022doltest}. 
\ding{183} On the other hand, an \textit{attacker-controlled user device} can also send malformed messages to the base station to which it connects \cite{kim2019touching, hussain20195greasoner}.
\ding{184} RRC messages \textit{transmitted between base stations} (e.g., during handovers) are also vulnerable. Studies \cite{sharwood2025korean, femtocell} show that femtocells (compact indoor base stations) can be compromised to exploit this interface.

\subsection{Challenges and Motivation of \system{}}
\label{subsec:challenges}

To motivate the design of \system{}, we analyze the unique challenges UPER poses for formal specification and verification, and reasons why existing verified parsing frameworks are insufficient for handling its complexities (Table~\ref{tab:verified_parser_comparsion}).

\cparagraph{Backward/Forward Compatibility with Extensions}
As discussed in \cref{sec:background}, ASN.1 extensions are fundamental to the evolution of 4G, 5G, and V2X protocols. 
This brings additional challenges because extensions would break the \textit{injection} property, as they require the parser to accept messages from different versions.
For example, in Figure~\ref{fig:bit_encoding}, when the old system receives the Message~\ref{subfig:example_msg_a} and attempts to re-encode it, the output is \texttt{0 0010 101}. This completely ignores all extensions (since none are present at Rel 15.2.1) and results in an encoding that differs from Figure~\ref{subfig:example_msg_big_encoding}.
Similarly, it is not a one-to-many relation between the value and the bit representation either, because while the parser is required to accept these unknown extensions, the serializer is strictly limited to the fields defined in its own version of the standard.

This makes adopting strict, non-malleable frameworks like Everparse \cite{Everparse19} or Comparse \cite{wallez2023comparse} difficult.
There are two potential workarounds, but they are both flawed. The first is to reject all messages containing unknown extensions, but this severely hinders interoperability in real networks where extensions are common. The second is to expand the input space to include all unknown extensions for the encoder, but this undermines the encoding model. 
Using a fully malleable format, such as Narcissus \cite{Narcissus19}, is also not appropriate since ASN.1 UPER is fundamentally not a malleable encoding. 
Thus, our approach will extend the round-trip theory.

\cparagraph{Bit-level Precision in a Majority Byte-Aligned World}
UPER is a bit-precise encoding format that operates at the level of individual bits rather than bytes, making implementation and formal verification more complex.
Decoding UPER requires maintaining a precise bit cursor and performing complex bitwise manipulation like shifting and masking to reconstruct parsed fields.

While it is possible to represent bit-level protocols in some systems like Narcissus \cite{Narcissus19}, their optimized extraction mechanisms only work with byte-aligned formats. Using these tools to handle bit-level granularity would incur significant performance penalties. 
Meanwhile, most other parser frameworks operate on byte-aligned formats, because common protocols like ASN.1 DER are byte-aligned. Thus, approaches like ASN1$^\star$~\cite{ni2023asn1} and EverParse~\cite{Everparse19} cannot be easily ported to UPER. 
This is because they use high-level parsers/serializers and low-level accessors/readers to track byte positions for zero-copy implementation. Adapting these systems to bit-level formats would require significant rewriting of both high- and low-level components to handle bit-level positions and may incur performance overhead. 
And their zero-copy feature offers minimal benefits for UPER since values at arbitrary bit positions cannot be dereferenced and still need bit-level extraction.

\cparagraph{Explicit and Implicit Constraints}
Constraints are essential for efficient serialization in UPER, but complicate verification. Explicit constraints, such as \textsf{INTEGER (1..8)} (Figure \ref{subfig:asn1_def_b}), enforce fixed value ranges that are easily formalized. On the other hand, implicit constraints during encoding add more complexity. For example, encoding an \textsf{INTEGER} requires converting the value to signed two's-complement binary, normalizing it to the most compact octet-aligned form, and prefixing it with a length determinant, while always using the minimal number of octets. 
Similar challenges arise with structural types like \textsf{SEQUENCE}, where fields of non-structural type with \textsf{DEFAULT} values must be omitted if their value is equal to the default.

\cparagraph{Encoding Open Type Fields}
Encoding ASN.1 Open Type Fields is nuanced. Introduced in \cref{sec:background}, these fields are prefixed with their length, which allows older parsers to skip unknown extensions (Figure~\ref{fig:bit_encoding}). This creates a circular dependency where the number of bits used for the length prefix itself depends on the total length of the value, which is only known after encoding is complete. Consequently, attempting to integrate these fields into verified parser frameworks like EverParse~\cite{Everparse19} is non-trivial, as it requires dynamic memory allocation and multi-pass encoding for each Open Type encounter.
Our approach calculates the length beforehand by keeping a separate length function, eliminating the need for a temporary buffer and copy operation.
However, this introduces additional complexity, as the length function must be verified to be consistent with the encoder to ensure correctness.
Furthermore, Open Type fields impose a special fragmentation process when the length exceeds 16K bytes. But this occurs very rarely in practice \cite{5gRRC_NR, LTE_RRC}, thus we limit the possible length for an Open Type encoding to 16K bytes (128K bits) for simplicity. 
While this limit is not an issue in practice, it has consequences for formal verification. 
Within the logic of the theorem prover, this constraint restricts the domain of possible messages for the parser and serializer, which complicates how we define certain correctness properties.

\begin{table}[h]
    \caption{Comparison with existing verified parsers}
    \label{tab:verified_parser_comparsion}

    \scriptsize
    \begin{tabular}{| l | l | c c c | c c c c |}
        \hline 
        \multirow{2}{*}{\textbf{System}} & \multirow{2}{*}{\textbf{Format}} &  \multicolumn{3}{c | }{\textbf{Properties}}  & \multicolumn{4}{c | }{\textbf{Features}} \\ 
        & & Surj & Inj & Weak-Inj &  Bit & Ext & Auto & Fast  \\ 
        \hline \hline 
        \textbf{\system{}} & ASN.1 UPER & \yes  &  \yes & \yes & \yes & \yes & \yes  & \no  \\  \hline 
        EverParse~\cite{Everparse19} & Custom DSL &  \yes & \yes & \no  & \no & - & \yes & \yes \\ \hline 
        Narcissus~\cite{Narcissus19} & Custom DSL &  \yes & \yes & \limited & \limited & - & \yes & \no \\ \hline 
        ASN1$^\star$~\cite{ni2023asn1} & ASN.1 DER & \yes & \yes & \no & \no & \no & \limited & \no  \\ \hline 
        ARMOR~\cite{debnath2024armor} & ASN.1 DER & \yes & \yes & \no & \no & \no  & \limited & \no  \\ \hline 
        V2V~\cite{tullsen2018cav} & ASN.1 UPER & \yes & \yes & \no & \yes & \no  & \yes & \yes \\ \hline 
        ASN1SCC~\cite{asn1acn25bucev} & ASN.1/ACN &  \yes & \no &\no  & \yes & \no & \yes & \no  \\  
        \hline     
    \end{tabular}
    { \yes{}: full support, \no{}: no support, \limited{}: limited support, - : not applicable;
    Ext: ASN.1 Extension; Auto: automatically generating a verified parser; Fast: efficient implementation.
    \textbf{Surj}ection; \textbf{Inj}ection }
\end{table}

\subsection{Design Philosophy}
\label{sec:verification_philosophy}

\begin{figure*}[ht]
    \centering
    \includegraphics[width=0.9\linewidth]{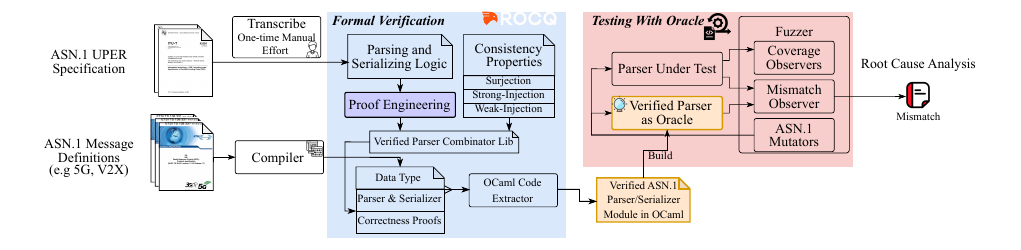}
    \caption{Architecture Overview}
    \label{fig:arch}
 
\end{figure*}

\para{Objective \& Scope}
Our objective is to formally verify the correctness of ASN.1 parser and serializer implementations. 
This practical focus on implementation distinguishes our work from efforts like ASN1$^\star$~\cite{ni2023asn1}, which develops a formal model of the ASN.1 language itself, treating verified DER parsers as executable semantics to ensure that any valid ASN.1 structure produces a verified parser. 
We therefore exclude a verified compiler: since ASN.1 semantics are defined by the encoders and decoders the compiler produces~\cite{steckler2007asn1formal}, verifying it is a tautological exercise of marginal value.

We use the Rocq prover~\cite{Coq-refman} to implement the verified parser and serializers. 
Termination is ensured by Rocq's type checker, as all recursive functions are defined by structural recursion over standard inductive types.
We do not explicitly prove memory safety, but rely on the memory safety provided by the OCaml runtime, and trust Rocq's extraction mechanism to preserve the properties established in the proof assistant.

\para{Extended Round-trip}
We choose to verify the parser’s correctness by proving an extended round-trip consistency with the corresponding serializer.
In contrast, previous approaches such as Narcissus~\cite{Narcissus19} and ARMOR~\cite{debnath2024armor} verify both the parser and the serializer against a single source of truth: a formal specification defined as a relation between a value and its encoding.
This approach offers advantages in settings like Protocol Buffers~\cite{ye19verified}, where one-to-many encodings must be supported. 
However, UPER does not involve one-to-many encodings as previously discussed (\cref{subsec:challenges}).
Moreover, these approaches ultimately provide mathematically equivalent guarantees but require additional proof overhead to show that the parser conforms to the bijective format.

\subsection{\system{} Workflow}

Figure \ref{fig:arch} illustrates the overall architecture of \system{}. The process starts with formalizing the concept of parser/serializer round-trip consistency properties (\cref{subsec:correctness_guarantees}). Next, we develop a set of verified basic combinators (\cref{subsec:basic_parser_comb}), which serve as the foundation for ASN.1 primitive types. Subsequently, we construct structural combinators (\cref{subsec:asn1_comb}), like \textsf{SEQUENCE}, and prove their correctness. We also implement a compiler (\cref{subsec:compiler}) that generates data types and verified parsers/serializers from any ASN.1 definition by leveraging our library of verified combinators.
One key benefit of a verified parser is that it can effectively work as a trusted \textit{test oracle} (\cref{sec:diff_testing}), enabling rigorous conformance testing of other parsers.

\section{Design Details of \system{}}
\label{sec:design_details}

We now discuss the design details of \system{}, including the parser/serializer definition, solution to the challenges, correctness guarantees, verified combinators and compiler. 

\subsection{\system{} Serializer and Parser Definitions}

Our framework is built on parsers and serializers that operate directly on a byte array $\mathbb{B}$ (where a byte is $\{0, \ldots, 255\}$), which we treat as a continuous bitstream. To navigate this stream, we use a precise bit position indicator, $\mathbb{P} := \mathbb{N} \times \{0,\ldots, 7\}$, representing a byte index and a bit offset within that byte.
Let $\textsf{Flg} := \textsf{SameVer} | \textsf{DiffVer}$ to indicate whether parsing results in the same version or a different version. 
To enforce type constraints, these parser/serializer functions are parameterized by a message type $A$ and a refinement function $P: A \to \textsf{Prop}$. This refinement maps a value of type $A$ to a logical proposition defining its constraint. For example, the constraint for \textsf{maxMIMO-Layers} in Figure \ref{subfig:asn1_def_b} is \textsf{INTEGER (1..8)}, so its refinement is $\textsf{fun } (x : \mathbb Z) \Rightarrow 1 \leq x \leq 8$. 
We define their type in the curried form as follows.
\begin{lstlisting}[style=mathstyle]
Parse $A$ $P$ := $\mathbb{B}$ $\to$ $\Pos$ $\to$ option($\{x : A | P x\}$ $\times$ $\Pos$ $\times$ Flg)
Serialize $A$ $P$ := $\mathbb{B}$ $\to$ $\Pos$ $\to$ $\forall$ $a$ : $A$, option($\B$ $\times$ $\Pos$ $\times$ $P$ $a$)
\end{lstlisting}
In essence, the parser takes a buffer and a starting bit position. Upon success, it returns a tuple containing: (1) the sigma type of the parsed value, along with a proof that it satisfies the condition $P$; (2) the new bit position after the read; and (3) the versioning flag. The serializer takes a buffer, a starting position, and a value to write. When successful, it returns the updated buffer, the new bit position along with a proof that the value satisfies the required condition $P$. 
Both functions are wrapped in an \textsf{option} type to handle errors, such as reaching the end of the buffer unexpectedly or encountering a value that violates its expected constraints.

\subsection{Solutions to Challenges of Verifying UPER}

\para{Weak-Injection Property}
To solve challenge \challengeref{1} (forward/backward compatibility), we adopt a flag-based version detection mechanism and a corresponding \textit{Weak-Injection} property~\cite{tan2018bidirectional}, formally defined in \cref{subsec:correctness_guarantees}. The parser sets a flag if it encounters an unexpected number of extensions, signaling a version mismatch. This lightweight approach extends security guarantees with minimal overhead.
Consequently, our correctness guarantees are two-tiered: the standard \textit{Injection} property applies when versions match, while the \textit{Weak-Injection} property applies in all cases. \textit{Weak-Injection} ensures that the serializer can handle any output from the parser, guaranteeing that even when re-encoding does not produce an identical bitstring, the resulting value remains valid and constrained rather than arbitrary.
The trade-off is the proof effort to establish this property for all extensible datatypes.

\para{Bitstream Approach} 
To better handle Challenge \challengeref{2} (bit-level precision), 
we utilize direct, efficient bit-level manipulation.
There can be some alternatives. 
One potential way is to map the input stream to a \textsf{list bool} in Rocq. While this approach seems easier in theory, it can be highly inefficient, 
especially when the ASN.1 messages can be several thousand bits long in complex network protocols (e.g., 4G, 5G). 
Another alternative involves encoding fields into independent byte buffers and subsequently concatenating them. However, this process is inefficient for nested structures, such as \textsf{SEQUENCE}, where each field might be realigned to maintain bit-level correctness and concatenated multiple times.
In contrast, \system{} operates directly on a bitstream while tracking the current bit offset, naturally supporting arbitrary bit alignment without requiring repeated shifting or copying.
We reformulate the consistency properties for our parser and serializer definitions without compromising the level of assurance, a non-trivial task discussed in \cref{subsec:correctness_guarantees}.

\para{Parser Combinators}
For Challenge \challengeref{3} (explicit \& implicit constraints), we use parser combinators to modularly construct data types, parser/serializers, and proofs, tackling ASN.1's complexities and enabling scalability of the verification process. For explicit constraints, we employ combinators like \textit{restrict} to layer value constraints over other parsers. 
For implicit constraints, we compose basic verified parser combinators, but the process itself is burdened with proofs.
To illustrate, supporting the \textsf{INTEGER} type requires defining a \textit{mapping} between the integer and a dependent pair of a length determinant, and the integer's two's complement form. The first task is to restrict the dependent pair to only the minimum representation. Next, we must prove that this mapping constitutes a bijection to maintain the \textit{round-trip} properties. This proof takes approximately 800 lines of code (LoC) because Rocq employs abstract mathematical representations of integers or natural numbers rather than machine integers, making seemingly simple procedures surprisingly complex. 
Finally, we design ASN.1-specific combinators to handle complex structural logic, as well as addressing \challengeref{2}. Details of combinators are in \cref{subsec:parser_combinators}.

\para{Open Type Field}
For Challenge~\challengeref{4}, we define a separate length function with type  $A\to \textsf{option }\mathbb{N}$, where the \textsf{option} type handles inputs violating specific constraints. 
Furthermore, restricting length of Open Types complicates \textit{weak-injection} property, as additional constraints are required to guarantee the production of valid encodings.

\subsection{Correctness Guarantees}
\label{subsec:correctness_guarantees}
\begin{figure}
    \centering
    \includegraphics[width=.9\linewidth]{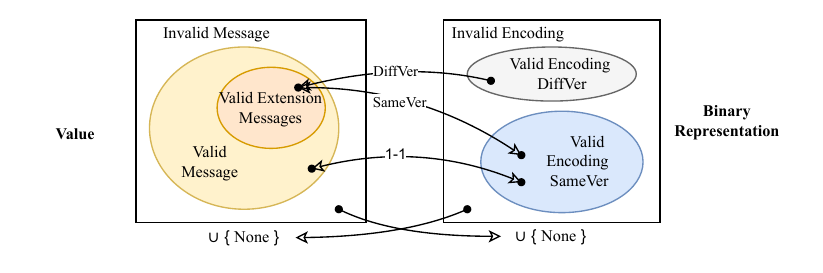}
    \caption{Mapping between values and their binary encoding.}
    \label{fig:property}
\end{figure}

Figure \ref{fig:property} illustrates the property we aim to prove.
We define the surjection and injection properties first. We assume that \lstinline[style=mathstyle] {$A$ : Set},
\lstinline[style=mathstyle]  {$P$ : $A$ $\to$ Prop}, 
\lstinline[style=mathstyle] {enc : Serialize $A$ $P$}, 
\lstinline[style=mathstyle] {dec : Parse $A$ $P$},
and the length helper function \lstinline[style=mathstyle] { msg_len : $A$ $\to$ option $\mathbb N$ },
Also, we assume \textit{proof irrelevance} in all definitions. The properties are summarized in Table \ref{tab:consist_prop}. 

\para{Surjection and Strong-Injection} 
We can directly embed the mathematical properties of surjection and injection (\cref{sec:high_level_parser_prop}) into the definitions of our serializer and parser.
For \textit{surjection}, we formally define the property as follows.
\begin{lstlisting}[style=mathstyle]
Definition surjection := $\forall$ ($b$ $b'$ : $\B$) ($p$ $p'$ : $\Pos$) $(a : A)$ (pf : $P$ $a$), 
  enc $b$ $p$ $a$ = Some ($b'$, $p'$, pf) $\to$
  dec $b'$ $p$ = Some (exist P $a$ pf, $p'$, SameVer).
\end{lstlisting}
This definition asserts that if encoding a value $a$ from position $p$ successfully yields a buffer $b'$ and position $p'$, decoding $b'$ from the same starting position $p$ must return the same $a$, and position $p'$.

Meanwhile, injection requires that if a buffer $b$ decodes to some value $a$, then re-encoding $a$ should yield the same buffer. 
However, equivalence between buffers must be carefully defined, as re-encoding creates a new buffer, with potentially different positions. To address this, we introduce the notion of \textsf{equiv\_bits}, which captures bitwise equivalence between two buffer slices. Its formal type is:
\lstinline[style=mathstyle] {equiv_bits ($b_1$ $b_2$ : $\B$) ($p_1$ $p_1'$ $p_2$ $p_2'$ : $\Pos$) : Prop}.
Intuitively, the slice of buffer $b_1$ in the window $[p_1, p_1')$ is identical, bit for bit, to the slice $[p_2, p_2')$ of buffer $b_2$. 
Additionally, \textsf{equiv\_bits} implies that $p_1 \leq p_1' \leq \textsf{buf\_len}(b_1)$ to ensure valid buffer bounds, where \textsf{buf\_len} returns the bit-length of the buffer.
Moreover, to avoid rejection due to insufficient buffer length, we quantify over any buffer with the same length as the original. We define \textit{strong injection}.
\begin{lstlisting}[style=mathstyle]
Definition strong_injection := $\forall$ ($b$ : $\B$) ($p$ $p'$ : $\Pos$) $(a : A)$ (pf : $P$ $a$), 
  dec $b$ $p$ = Some (exist P $a$ pf, $p'$, SameVer) $\to$
  $\forall$ ($b_0$ : $\B$), buf_len $b_0$ = buf_len $b$ $\to$
    $\exists$ ($b_1$ : $\B$), enc $b_0$ $p$ $a$ = Some ($b_1$, $p'$, pf) $\land$ equiv_bits $b$ $b_1$ $p$ $p'$ $p$ $p'$.
\end{lstlisting}
However, such straightforward transcription would not suffice. The defined \textit{strong-injection} properties do not imply \textit{non-malleability}.
\begin{lstlisting}[style=mathstyle]
Definition non_malleability := 
$\forall$ ($b_1$ $b_2$ : $\B$) ($p_1$ $p_1'$ $p_2$ $p_2'$ : $\Pos$) ($a$ : $A$) (pf : $P$ $a$)
  dec $b_1$ $p_1$ = Some (exist $P$ $a$ pf, $p_1'$, SameVer) $\to$
  dec $b_2$ $p_2$ = Some (exist $P$ $a$ pf, $p_2'$, SameVer) $\to$
    equiv_bits $b_1$ $b_2$ $p_1$ $p_1'$ $p_2$ $p_2'$.
\end{lstlisting}
This is due to the potential dependency of parsers on external factors like bit position. 
For instance, consider two non-malleable serializer-parser pairs, $(\textsf{enc}_1, \textsf{dec}_1)$ and $(\textsf{enc}_2, \textsf{dec}_2)$. 
We can construct a third serializer-parser pair that uses $(\textsf{enc}_1, \textsf{dec}_1)$ if the bit position $\leq (1, 0)$ and uses $(\textsf{enc}_2, \textsf{dec}_2)$ otherwise. The third pair satisfies this version of \textit{strong injection} property, yet it clearly fails to produce a unique encoding.
Therefore, we need to ensure the encoder is independent of both the given buffer and write position. 

\begin{lstlisting}[style=mathstyle]
Definition encode_invariance :=
$\forall$ ($b_1$ $b_1'$ $b_2$ $b_2'$ : $\B$) ($p_1$ $p_1'$ $p_2$ $p_2'$ : $\Pos$) ($a$ : $A$) (pf : $P$ $a$), 
  enc $b_1$ $p_1$ $a$ = Some ($b_1'$, $p_1'$, pf) $\to$ enc $b_2$ $p_2$ $a$ = Some ($b_2'$, $p_2'$, pf) $\to$ 
  equiv_bits $b_1'$ $b_2'$ $p_1$ $p_1'$ $p_2$ $p_2'$.
\end{lstlisting} 

\noindent This property ensures the serializer produces a canonical bit-level representation for any value, preventing context-dependent outputs.
Thus, we can prove that a combination of \textit{strong-injection} and \textit{encode-invariance} implies \textit{non-malleability}. 
Similarly, \textit{surjection} does not imply \textit{non-ambiguity} automatically. We define the dual property \textit{decode‐invariance}, to ensure that the parser is not dependent on unrelated factors like position or buffer content.

\para{Local Write} One other important property is \textit{local write}. It ensures that encoding into a buffer \( b \) from position \( p \) to \( p' \) does not modify bits outside this range. Here, \( \bm{0} := (0,0) \) indicates the starting position of a buffer, while \( b.\mathbf{end} \) indicates the end. 
\begin{lstlisting}[style=mathstyle]
Definition local_write := $\forall$ ($b$ $b'$ : $\B$) ($p$ $p'$ : $\Pos$) $(a : A)$ (pf : $P$ $a$), 
  enc $b$ $p$ $a$ = Some ($b'$, $p'$, pf) $\to$
  equiv_bits $b$ $b'$ $\bm{0}$ $p$ $\bm 0$ $p$ $\land$ equiv_bits $b$ $b'$ $p'$ $b.\textbf{end}$ $p'$ $b'.\textbf{end}$ $\land$ $p$ $\leq$ $p'$.
\end{lstlisting}
This property is crucial for composing serializers and parsers, such as in a format that concatenates encoded types $A$ and $B$.
This property prevents the serialization of $B$ from corrupting a previously-written bit segments of $A$. 
With it, we can confidently chain proofs, applying $A$'s surjection proof to its segment without interference from $B$.
Additionally, \textit{local write} implies that the buffer length remains unchanged between $b$ and $b'$, a useful implication in establishing \textit{strong-injection}.

\para{Weak-Injection Property} 
This property aims to prevent malformed packets while maintaining flexibility.

\begin{lstlisting}[style=mathstyle]
Definition weak_injection :=
  $\forall$ ($b$ : $\B$) ($p$ $p'$ : $\Pos$) $(a : A)$ (pf : $P$ $a$) (flg : Flg), 
    dec $b$ $p$ = Some (exist P $a$ pf, $p'$, flg) $\to$
    $\exists$ $l_{\min}$ : nat, $\forall$ ($p_0$ : $\Pos$) ($b_0$ : $\B$), 
      to_nat $p_0$ + $l_{\min}$ $\leq$ buf_len $b_0$  $\to$  msg_len a $\leq'$ 128K $\to$
      $\exists$ ($b_0'$ : $\B$) ($p_0'$ : $\Pos$), enc $b_0$ $p_0$ $a$ = Some ($b_0'$, $p_0'$, pf) $\land$
        to_nat $p_0'$ $\leq$ to_nat $p_0$ $+$ $l_{\min}$.
\end{lstlisting}
The definition establishes that if we can decode some buffer $b$ to value $a$, then $a$ can be encoded into some buffer $b'$, provided sufficient space. 
An additional constraint applies to the length of an encoding of $a$, without which, the length limit on Open Type field cannot guarantee the success of the encoding procedure at the end.

Finally, the \textit{length-consistency} property ensures the length function \textsf{msg\_len} accurately computes the encoding length as the difference between final and initial positions. This answers the Open Type field issue in Challenge \challengeref{4}, and aids the proofs of properties like \textit{surjection} for Open Type fields.

\begin{table}[h]
    \caption{Correctness properties}
    \label{tab:consist_prop}
    \centering
    \scriptsize
    \begin{tabular}{ | l | p{5.8cm} | }
         \hline 
        \textbf{Property} & \textbf{Description} \\
        \hline
\textit{Surjection} & 
Preventing miscommunication
\\  
    \textit{Strong-Injection} & 
Rejecting all non-canonical packets
\\ 
\textit{Weak-Injection} & 
Prevent malformed values 
\\ 
\hline 
        \textit{Local Write} 
&  Preventing unintended writes to the buffer
\\  
    \textit{Encode-Invariance} 
& Ensuring the serializer is buffer or position-independent.
 \\ 
    \textit{Decode-Invariance} 
& Ensuring the parser is buffer- or position-independent.  
 \\ 
\textit{Length Consistency} 
& Crucial for verifying Open Type Field serializer correctness.
\\ \hline   
    \end{tabular}
\end{table}

\subsection{Verified Parser Combinators}
\label{subsec:parser_combinators}

With correctness guarantees defined, we next discuss our combinator construction and proof engineering processes. We first introduce a notation for \textsf{Fmt ($A$: Set) ($P$ : $A$ $\to$ Prop)} as a record that encapsulates the parser, serializer, and associated round-trip consistency proofs 
(\extref{sec:additional_definitions}). This restructuring does not affect semantics or guarantees but improves clarity and facilitates combinator use.

\subsubsection{Basic Parser Combinators}
\label{subsec:basic_parser_comb}

\para{Basic Bit $n$ Format}
While not strictly a combinator, it acts as the direct interface function between the byte array $\B$ and the data type \textsf{nat}, enabling basic bit-level read/write operations and works as the cornerstone of all other parsers and serializers. 
The decoder loads a byte sequence into a native integer to extract bit-fields using bitwise shift and mask operations.
The encoder employs a read-modify-write process to maintain \textit{local write}, ensuring unrelated bits are not accidentally modified. The target bytes are loaded into an integer, modified only at the precise bit offset, and then written back to the buffer.
These proofs are complex, requiring precise, low-level reasoning about bitwise manipulations and bijective bytes-to-integer conversions.

\para{Basic Combinators} 
To construct parsers for basic types like \textsf{BOOLEAN} or \textsf{OCTET STRING} from our primitive bit parser, we use a set of format combinators.
The \textit{map} combinator, for instance, transforms an existing format into a new data type with the same binary representation. Consequently, a \textsf{BOOLEAN} format can be efficiently modeled using the \textit{map} and a basic bit format. 
Basic combinators can also propagate the metadata Flag.

\begin{table}[h]
    \caption{Selected combinators (refinement omitted)}
    \label{tab:format_comb}
\footnotesize
    \centering
    \scriptsize
    \begin{tabular}{| p{1.2cm} | p{2.5cm} |  p{3.5cm} |}
     \hline 
        \textbf{Combinator} & \textbf{Input (Correctness Rules)} $\to$ \textbf{Output} & \textbf{Description}  \\
        \hline \hline 
        \textit{basic-bit} $n$ &  
        \textsf{Fmt} $\mathbb N$  & 
        Formats a natural number between $[0, 2^n)$ as an n-bit integer.\\ 
        \hline 
        \textit{map} & 
        $s_A :$ \textsf{Fmt} $A$, and bijective map between $A$ and $B$ $\to$ \textsf{Fmt} $B$
        & 
        Formats type $B$ by applying the bijective function to the existing format of $A$ \\ \hline 
        \textit{dependent append} & 
        $s_A :$ \textsf{Fmt} $A$ , and $s_B :$ $A \to $ \textsf{Fmt} $B$ $\to$ \textsf{Fmt} $(A \times B)$ &  
        Formats a pair $(a, b)$ where the format for the second element $b$ depends on the value of the first element $a$ \\ \hline
        \textit{list} & 
        $s_A :$ \textsf{Fmt} $A$,
        $n : \mathbb N$ $\to$ \textsf{Fmt} (\textsf{list} $A$) &
        Formats a fixed-length list of $n$ elements, each of type $A$.
        \\ \hline  
        \textsf{CHOICE} & 
        $\forall i \in 1..n$, $s_i :$ \textsf{Fmt} $A_i$  $\to$ \textsf{Fmt} (\textsf{nth} $A_i$) & 
        Formats a tagged union that represents a value that can be one of several predefined types
        \\ \hline 
        \textsf{SEQUENCE} & 
        $\forall i \in 1..n$, $s_i :$ \textsf{Fmt} $A_i$, 
        $\textsf{mod}_i \in \textsf{\{Opt, Dft, Nor\}}$ $\to$ 
        \textsf{Fmt} $(A_1' \times \ldots(A_n'\times \top))$ & 
        Formats an ordered collection of items of different types, with modifiers for each item \\ \hline 
            
    \end{tabular}
\end{table}

\subsubsection{ASN.1 Combinators}
\label{subsec:asn1_comb}
While general-purpose combinators are versatile enough to model ASN.1 combinator types like \textsf{SEQUENCE-OF} via \textit{dependent append} and \textit{list} format, they struggle with the intricate requirements of extensible \textsf{SEQUENCE} types.
First, constructing a \textsf{SEQUENCE} using generic combinators like \textit{dependent-pair} and \textit{map} leads to verbose definitions, especially when handling inter-field dependencies, such as the optional-field bitmap.
More importantly, the basic parser combinators are designed to be bijective by default, but the \textsf{SEQUENCE} parser needs to identify and skip unknown extensions. 
It is natural to leverage the structured nature of ASN.1 by introducing a set of specialized, ASN.1 format combinators for types such as \textsf{CHOICE} and \textsf{SEQUENCE}, which correctly handle their complex extension mechanisms.

\para{Sequence} 
The \textsf{SEQUENCE} combinator maps an ordered list of fields, marked as normal, \textsf{OPTIONAL}, or \textsf{DEFAULT}, to a right-nested tuple (shown in Table \ref{tab:format_comb}). Field types are adjusted based on their modifiers, e.g., converting OPTIONAL fields to option types.
However, proving round-trip consistency is not straightforward because the \textsf{OPTIONAL} field bitmap creates a non-local dependency between presence bits at the start of the \textsf{SEQUENCE} and field data later in the stream.
Our solution is to prove the components separately. First, we prove the round-trip consistency of the bitmap. 
Second, we inductively prove field consistency by parameterizing the parser with the decoded bitmap.
Finally, we use a prepend correctness property to formally link these two separate proofs, guaranteeing the correctness of the entire \textsf{SEQUENCE} encoding.

\para{Sequence Extensions}
Parsing extensions is complicated due to the potential mismatch between the number of decoded extensions ($n$) and the number expected by the current version ($m$). 
While \textsf{weak-injection} guarantees a valid encoding for the length of $n$, it fails to establish the validity of the count $m$'s encoding.
We thus define a valid-encoding property ensuring that any value satisfying a specific refinement condition has a valid encoding. We prove this property for the length determinant and the \textsf{list} combinator to validate the encoding of length $m$ and the extension bitmap.

\subsection{Compiler}
\label{subsec:compiler}

To handle the scale of real-world protocols like 5G, we develop an automated compiler that translates ASN.1 definitions into verified Rocq code.  
The compiler recursively generates types and formats using the verified combinators. It also aims to provide an intuitive and user-friendly interface, for example, mapping \textsf{SEQUENCE} to a \textsf{Record} type instead of a nested product. 
The compiler also generates helpful lemmas and proof obligations. For example, it may establish a one-to-one mapping between an \textsf{ENUMERATED} type and a constrained integer with the proofs generated using custom tactics.
More details are discussed in \extref{subsec:compiler_workflow}.

\section{Testing with Verified Oracle} \label{sec:diff_testing}

\system{}’s formal guarantees make it a unique trusted oracle for testing. 
By comparing other parsers against it, we can detect subtle implementation flaws. This approach is particularly effective at uncovering bugs common to all implementations, which is a critical blind spot for traditional differential testing~\cite{oracle_problem}. 
This process also serves a dual purpose as it provides crucial empirical validation for \system{} itself. Formal verification is only as strong as its underlying specification. By testing against widely-used systems, we confirm that our interpretation of the natural-language ASN.1 standard is correct and ensure real-world interoperability. Any discrepancies can indicate a non-conformance in the tested program, but also provide a feedback loop for refining our implementation.

\para{Distilling Executable Parser} 
We extract the verified parser into OCaml via extraction in Rocq. Our Trusted Computing Base (TCB) includes Rocq (and its extraction process), the OCaml compiler, and the OCaml Bigarray library, which implements our axioms for byte buffers. 
For practical performance during fuzzing, we directly map Rocq's arbitrary-precision \textsf{nat} type to OCaml's machine integer. To mitigate overflow risks and ensure the extracted code remains consistent with our formal model, we cap the \textit{basic-bit $n$} format at 48 bits. 
While using tools like CertiCoq~\cite{anand2017certicoq} could reduce the TCB, it can incur more overhead and deviate from our primary focus on parsing logic. We consider our approach sufficiently reliable.

\para{Structure-Aware Fuzzing}
We utilize AFL++ \cite{aflplusplus} as our primary fuzzing engine, while employing LibAFL \cite{libafl} specifically for baseband testing due to its modularity and seamless integration with our emulation environment. 
To navigate the complexities of ASN.1, we implement structure-aware fuzzing by incorporating ASNFuzzGen \cite{ransack24}, which can linearize ASN.1 messages to an internal byte-level abstraction. 
This custom mutator starts by linearizing the ASN.1 message, then mutates the resulting byte buffer, and finally delinearizes the byte to reconstruct the final ASN.1 message.
By combining this structure-aware approach with traditional havoc mutation, we can explore novel message instances that remain structurally plausible yet purposefully deviate from the specification.

\para{Output Normalization}
To properly compare parsers beyond a simple accept/reject check, we must compare their output values. 
However, it is challenging due to the different internal representations of data (C uses structs; Python uses dictionaries). 
To address this, we first normalize all outputs to a standardized 3GPP format using the JSON Encoding Rules (JER)~\cite{ITU_X697} as a middle ground. 
For our own verified parser, we automate this JER serialization using the \texttt{ppx\_deriving} preprocessor. 
For the parser-under-test, all have implemented the JSON encoder as part of their toolchain, with the exception of baseband targets. 

For these targets where reverse engineering internal data formats is impractical, we instead use a simpler metric: the number of bits each parser consumes. This metric is always available, since any ASN.1 UPER parser must accurately track bit positions to decode values correctly.
\Extref{sec:additional_background} further illustrates JER output normalization with the example in Figure~\ref{subfig:example_msg_a}. The only caveat is that the reference and target parsers may occasionally reject a message for different internal reasons. But we consider this case out of scope, since both parsers can reject this message.

\section{Implementation Details} \label{sec:implementation}
We estimate developing the 12k lines of code (LoC) Rocq framework and its 1.8k LoC OCaml compiler takes 4-5 person-months, as it requires significant manual proofing despite automation tactics.
The varied internal data representations across ASN.1 libraries force us to write custom harnesses for each differential testing target. 
For closed-source baseband targets, we used Ghidra~\cite{ghidra} to identify relevant parsing functions and Unicorn~\cite{unicorn} for emulation. 

\para{Supported Features} We support ASN.1 structures, including \textsf{SEQUENCE}, \textsf{CHOICE}, and \textsf{SEQ-OF}, all with extensions. Terminal types include \textsf{BOOLEAN}, \textsf{INTEGER}, \textsf{ENUMERATED}, \textsf{BIT/OCTET STRING}, and special strings including \textsf{UTF8String} and \textsf{IA5String}.  This also includes features used by the constructs above, including extensions in terminal types, length determinant (Normally Small Length Determinant), and Open Type Fields.  
We omit encoding rules \textsf{SET-OF} and fragmentation that are barely used in practice. 
We also omit certain high-level language abstractions, including \textsf{CLASS}, as they do not appear in our target protocols and do not impact the message encoding. 

\section{Evaluation} \label{sec:evaluation}

We experimentally evaluate our verified parser for its applicability, correctness, and performance with the following key questions:
\begin{enumerate}[label=\textbf{Q\arabic*.}]  
\item How accurate and precise is the {\system} parser compared to other ASN.1 implementations? (\cref{subsec:vuper_accuracy})
\item How effective is \system{} in identifying non-compliance using our differential testing framework? (\cref{subsec:identified_issues})
\item What are the consequences of the non-compliances? (\cref{subsec:attacks})
\item How performant is the verified parser compared with other ASN.1 parsers? (\extref{subsec:performance})
\item Baseline Comparison. How does \system{} compare to negative testing and consensus testing? 
\end{enumerate}

\subsection{Evaluation Setup}

\para{Targeted Protocols} 
To ensure practical relevance, we evaluate our parser on messages from the most impactful usecases of ASN.1 UPER: mobile and vehicular communication. We select the following representative message types for our evaluation:

\begin{description}
    \item[\textit{RRC Messages in 4G/5G.}] We target complex control-plane messages (\textsf{DL/UL-DCCH-Message}) used for critical signaling, such as handovers, between the User Equipment (UE) and base station~\cite{5gRRC_NR, LTE_RRC}. They contain more than 800 submessages and utilize diverse ASN.1 structures. 

    \item[\textit{V2X messages.}] We test key vehicle-to-everything (V2X) messages from the ETSI ITS standard~\cite{CAM_ITS, DENM_ITS}. Cooperative Awareness Messages (CAMs) periodically broadcast a vehicle's state (e.g., position, speed), while event-driven Decentralized Environmental Notification Messages (DENMs) warn of specific hazards like accidents. While smaller, the V2X messages introduce distinct elements not found in RRC messages, such as extended \textsf{INTEGER}.
    
\end{description}

\para{Targeted ASN.1 Systems}
Our evaluation includes 7 open-source tools: lionet asn1c~\cite{ASN1c}, pycrate~\cite{pycrate}, asn1tools~\cite{asn1tools}, TITAN~\cite{titan_asn1}, two parsers from srsRAN~\cite{srsRAN,srsRAN_Project} and rasn~\cite{rasn}; and 4 proprietary ones: Objective Systems' ASN1C~\cite{objSysASN1C} (denoted obj-sys), two versions of ProASN (parsers extracted from Samsung S21 and Google Pixel 6 basebands), and ffasn1 \cite{ffasn1}. 
Some are general-purpose ASN.1 parser generators; commercial basebands and srsRAN only work for specific 4G/5G message types; and ffasn1 is a blackbox message dump. 
Feature support also varies: asn1tools lacks support for parameterized types used in 5G, so we only test it on V2X messages. 

\para{Methodology} We employ a dual-input evaluation strategy to validate both the practical utility and the capability of finding non-compliance using \system{}. 
First, we test our verified parser against a real-world dataset of 5G RRC traces captured from live UE-base station communications to demonstrate conformance. 
Second, we evaluate the ASN.1 parsers-under-test using fuzzing with a verified oracle, as discussed in \cref{sec:diff_testing}. 
Our fuzzing experiment is done using Intel(R) Xeon(R) Gold 6448H @ 2.4GHz CPUs, and we do 5 runs on one specific message (like 5G DL-DCCH-Message) for 24 hours. 
Additionally, to minimize false positives arising from specification version discrepancies, we generate all open-source parsers from the same ASN.1 source. While newer releases often utilize extensions for compatibility, they occasionally alter the message structure via mandatory field changes. This remains a concern in proprietary baseband parsers under test since their specific implementation versions are unknown.

\subsection{\system{} Accuracy (Q1)} 
\label{subsec:vuper_accuracy}
We first evaluate \system{} with two sets of datasets: (1) real-world network operators' and user devices' downlink (DL) and uplink (UL) traffic, and (2) traffic generated through fuzzing.  
And we look for the following aspects for accuracy: overall accept/reject of a message, compatibility, JER encoding, and False Positives. 

\para{Real-World Dataset}
Our comparison on real-world data yielded two primary observations, as shown in Table \ref{tab:real_world_data}. 
First, all baseline parsers accepted every collected message without exception. Second, \system{} accepted the majority of messages but rejected a specific subset. For the messages accepted by both \system{} and the baseline targets, we find no discrepancies in the parsed results. Further investigation into the messages rejected by \system{} reveals that all of them contained explicitly encoded \textsf{DEFAULT} values. Per X.691~\cite{ITU_X691}, `encodings of components marked \textsf{DEFAULT} shall always be absent if the value to be encoded is the default value of a simple type'. This indicates that the encoders used by the specific network operator are non-compliant with the standard, a deviation that \system{} correctly identifies while other parsers overlook. 

\begin{table}[h]
    \caption{Real-world 5G data reject-accept comparison}
    \label{tab:real_world_data}
    \centering
    \scriptsize
    \begin{tabular}{ | l | l | l | l | }
         \hline 
        \textbf{Message Source} & \# Msgs & \system{} (Acc-Rej)  & Baseline (Acc-Rej)  \\
        \hline 
        \textbf{Country-1 Operator} (5G DL)  & 16644 &  16644 - 0 &  16644 - 0  \\ \hline 
        \textbf{Country-1 Operator} (4G DL) & 47689 & 47689 - 0 &  47689 - 0 \\ \hline 
        \textbf{Country-2 Operator} (5G DL)  & 18499 & 14593 - 3906 &   18499 - 0 \\  \hline 
        \textbf{Qualcomm baseband} (5G UL) & 42596 &  42596 - 0 & 42596 - 0 \\ \hline 
        \textbf{Qualcomm baseband} (4G UL) & 60970 & 60970 - 0 &  60970 - 0
        \\ \hline 
    \end{tabular}
\end{table}

\para{Fuzzer Generated Dataset} We summarize the accuracy on the fuzzer-generated traffic for four parsers in Table~\ref{tab:fuzzing_data}. Over 99\% of generated inputs produced no discrepancies, confirming that \system{} is largely conformant with standard implementations. 
To systematically triage the remaining discrepancies, we apply a diagnostic pipeline that categorizes behavioral differences into three classes: (I) \system{} Rejection $V_{\text{rej}}$ and Target Acceptance $T_{\text{acc}}$;
(II) \system{} Acceptance $V_{\text{acc}}$ and Target Rejection $T_{\text{rej}}$; and (III) Dual Acceptance and Semantic Divergence. 
Class I typically corresponds to injection violations, and Class II/III to surjection violations. 
However, Class I discrepancy can also result from surjection violations where the target parser accepts a message not because a specific constraint is missing, but because a prior error caused the bit-cursor to shift. 

\para{Forward/Backward Compatibility Test} 
To evaluate the forward/backward compatibility of \system{}, we analyzed its parsing results across different protocol evolutions (5G RRC Release 16 and Release 17) and 
observed two distinct categories of schema changes:
(i) \textit{Standard Extension Changes:} Expected modifications utilizing standard ASN.1 extension markers; and 
(ii) \textit{Structural Redefinitions:} Alterations to existing message types (e.g., modifying a message definition from an empty \textsf{SEQUENCE} to one containing populated fields), found in 7 message types (\extref{sec:additional_compatibility_experiment}). 
Redefinitions inherently break backward compatibility and depart from ASN.1 evolution principles, so we filter out such messages.

\begin{table}[h]
    \caption{Forward Compatibility Tests (Inputs with Explicit Extensions)}
    \label{tab:forward_comp}
    \centering
    \scriptsize
        \centering
        \begin{tabular}{ | l | l | l | l | }
              \hline 
             \diagcell{Rel-16}{Rel-17} & Accept-SameVer & Accept-DiffVer & Reject  \\ \hline 
            Accept-SameVer & 0 & 0  & 0   \\ \hline 
            Accept-DiffVer & 25241 &  0 & 1623  \\ \hline 
            Reject  & 0  & 0  & 1952 \\
            \hline  
        \end{tabular}
    {\footnotesize \\ Rows and columns represent Rel-16 and Rel-17 outcomes, respectively.
    For example, location (Rel-16 Accept-DiffVer, Rel-17 Accept-SameVer) represents the number of messages (25241) accepted by Rel-16 with flag DiffVer and Rel-17 with SameVer. }
\end{table}
First, to evaluate \system{}’s forward compatibility and handling of unknown extensions, we generate 28816 Rel-17 messages via ASNFuzzGen~\cite{ransack24} and filter them using pycrate~\cite{pycrate} to isolate inputs with populated extension fields. We then evaluate this corpus against both Rel-16 and Rel-17 parsers. 
Table~\ref{tab:forward_comp} presents the results: most messages are accepted by both implementations, with the Rel-16 parser appropriately marking \textsf{DiffVer} (i.e., different version/release) to denote unknown extensions, 
thereby confirming forward compatibility. 
\Extref{sec:additional_compatibility_experiment} provides an example message showing the field changes. 
We note two other expected cases: 
(1) mutually rejected from fuzzer-generated malformed inputs; (2) rejected from Rel-17 but accepted by Rel-16. In the latter, the malformed data sits within a new extension that the Rel-16 parser correctly bypasses as unknown. But the Rel-17 parser recognizes the field and triggers a decoding failure.

Second, we evaluate the backward compatibility by feeding a total of 667825 inputs generated by ASNFuzzGen of Rel-16 to both parsers (filtered to remove structural redefinitions). Presented in Table~\ref{tab:backward_comp}, this result shows that the Rel-16 parser accepts most inputs as \textsf{SameVer} (i.e., the same version/release), while the Rel-17 parser appropriately flags 21k inputs as \textsf{DiffVer}. The only exception was a specific subset of messages flagged as \textsf{DiffVer} by both implementations. This is because those inputs contain the extension fields undefined by both the Rel-16 and Rel-17 specifications, prompting both parsers to identify a version mismatch. 

\begin{table}[h]
    \caption{Backward Compatibility Tests (Rel-16 Inputs from Fuzzer)}
    \label{tab:backward_comp}
    \centering
    \scriptsize
        \centering
        \begin{tabular}{ | l | l | l | l | }
              \hline 
             \diagcell{Rel-16}{Rel-17} & Accept-SameVer & Accept-DiffVer & Reject  \\ \hline 
            Accept-SameVer & 566701 & 21175 & 0  \\ \hline 
            Accept-DiffVer & 0 & 33 & 0   \\ \hline 
            Reject & 0 & 0 & 79916 \\
            \hline 
        \end{tabular}
    
\end{table}

\para{Root Cause Analysis} Our root cause analysis strategy is as follows: 
First, for Class I errors, we retrofit our verified OCaml parser to \textit{emit semantic error codes}. This allowed us to aggregate thousands of different discrepancies into a set of root causes. 
Second, for Class II and III discrepancies, we perform manual triage by \textit{dissecting the bit-packets}, supplemented by pycrate~\cite{pycrate} to visualize the field offsets.
Finally, to confirm a suspected root cause, we conduct \textit{targeted case studies} by injecting a single type of non-compliance, like an out-of-range integer, into a minimal ASN.1 definition or specific protocol messages. This approach reveals how the parser handles individual constraints without the noise of multiple errors.  
If possible we also conduct source code inspection.

\begin{table}[h]
    \caption{Fuzzing statistics for \textsf{DL-DCCH} message}
    \label{tab:fuzzing_data}
    \centering
    \scriptsize
    \begin{tabular}{| l | l | l | l | l | l |}
         \hline 
        \textbf{Category} & \textbf{Root Cause} & \textbf{asn1c} & \textbf{pycrate} & \textbf{ProASN} & \textbf{Obj-Sys} \\ \hline 
        Inputs Generated (million) &  - & 8m & 3.3m & 1.7m & 16m \\ 
        \hline 
        Saved Discrepancies & - & 736 & 1203 & 112 & 1682 \\ \hline
        \multirow{4}{*}{I ($V_{\text{rej}}$ \& $T_{\text{acc}}$)} & Total & 590 & 1180 & 85 &  1674 \\  
            & \hyperlink{W1}{W1} & 7 & 0 & 0 & 0 \\ 
            & \hyperlink{S2}{S2} & 38 & 80 & 14 & 96 \\ 
            & \hyperlink{S4}{S4} & 0 & 123 & 1 & 207 \\ \hline
         \multirow{2}{*}{II ($V_{\text{acc}}$ \& $T_{\text{rej}}$)} & Total & 71 & 11 & 7 &  0 \\  
         & R1 & 71 & 0 & 0 & 0\\ \hline 
         III ($V_{\text{acc}}$ \& $T_{\text{acc}}$ Diff) & R5 & 0 & 0 & 16 & 0 \\ \hline 
        False Positive & - & 75 & 12 & 4 & 8  \\ \hline 
        False Positive Rate & - & 10.2\% & 1.0\% & 3.6\% & 0.5\% \\ \hline
    \end{tabular}
\end{table}
\para{False Positives}
We find a class of false positives stemming from the padding bits within the Open Type field. 
While an Open Type length is specified in octets, the encapsulated data is bit-encoded. This discrepancy can result in 1–7 redundant bits at the end of the final octet. In the verified parser, to ensure full canonicality, we enforce all the redundant bits to be 0. Because these bits are never decoded into concrete values, the standard does not explicitly require them to be 0. Therefore, we do not consider parsers that omit this check to be inherently non-compliant.

\para{JER Accuracy}
There is no guarantee of correctness for the native JER encoders across the tested implementations; a flawed encoder could emit an incorrect JSON representation, artificially inflating the discrepancy rate. We resolved this through root cause analysis to eliminate such false positives.
For instance, during our analysis, we discovered and fixed a bug within the srsRAN~\cite{srsRAN_Project} JER encoder that caused it to emit invalid empty JSON objects. 
To ensure fairness, we normalize the extension group representation as we compare \system{} with different parser-under-test.

\subsection{Identified Issues via \system{} as Oracle (Q2)} 
\label{subsec:identified_issues}

\begin{table*}[t]
    \caption{Non-compliance with X.691 standard}
    \label{tab:violation}
    \scriptsize
    \begin{center}
    \begin{tabular}{ | l | l | l | l | l | l  | }
        \hline 
        \textbf{Flaw} & {\textbf{Non-Compliance}} & \textbf{X.691} & \textbf{Type} & \textbf{Security Impact} & \textbf{Impacted Implementations}  \\
        \hline 
        \hypertarget{W1}{W1} & Missing \textsf{INTEGER} Constraint Check & 11.5 & Weak-Inj & Vul (DoS, MC) & asn1c, asn1tools, rasn \\
        \hline 
        \hypertarget{W2}{W2} & Missing \textsf{SIZE} Constraint Check & 11.9.4.1 & Weak-Inj & Vul (DoS, MC) & asn1c, asn1tools, rasn \\
        \hline 
        \hypertarget{R1}{R1} & Limited \textsf{SEQ-OF} length & 20 & Surj-1 & NC & asn1c \\ \hline 
        \hypertarget{R2}{R2} & Incorrectly Rejecting more than 64 exts & 19.8 & Surj-1 & SD & pycrate \\ \hline 
        \hypertarget{R3}{R3} & Error with \textsf{UTF8String} Constraints & 30.6 & Surj-2 & SD & pycrate \\ \hline 
        \hypertarget{R4}{R4} & Incorrect Handling of Unknown Ext & 19 & Surj-2 & Vul (DoS) & srsRAN \\ \hline 
        \hypertarget{R5}{R5} & Incorrect Handling of more than 64 exts & 19.8 & Surj-2 & SD & ProASN \\ \hline 
        \hypertarget{R6}{R6} & Incorrect Handling of Normally Small Length Det & 11.9.3.4 & Surj-2 & SD & rasn \\ \hline  
        \hypertarget{S1}{S1} & Overlong Encoding (\textsf{INTEGER})$^\dagger$ & 11.4.6 & Strong-Inj & NC & asn1c, asn1tools, pycrate, rasn, ffasn1 \\ \hline 
        \hypertarget{S2}{S2} & Overlong Encoding (Length Determinant) & 11.9.3.6 & Strong-Inj & NC & asn1c, asn1tools, pycrate, srsRAN, ProASN, TITAN, obj-sys, rasn, ffasn1 \\  \hline 
        S3 & Overlong Encoding (Normally Small Length Det) & 11.9.3.6 & Strong-Inj & NC & asn1c, asn1tools, pycrate, srsRAN, ProASN, TITAN, obj-sys, ffasn1 \\  \hline 
        \hypertarget{S4}{S4} & Missing Open Type length validation & 11.2.2 & Strong-Inj & SD & asn1tools, pycrate, srsRAN, ProASN, TITAN, obj-sys, rasn, ffasn1 \\
        \hline 
        S5 & Accepting Extensions all absent & 19.9 & Strong-Inj & NC & asn1c, asn1tools, pycrate, srsRAN, ProASN, TITAN, obj-sys, rasn, ffasn1 \\
        \hline 
        S6 & Accepting ExtensionAdditionGroup all absent & 19 & Strong-Inj & NC & asn1c, asn1tools, pycrate, srsRAN, ProASN, TITAN, obj-sys, rasn, ffasn1 \\
        \hline 
        S7 & Accepting Encoding of Default Value & 19.5 & Strong-Inj & NC  & asn1c, asn1tools, pycrate, srsRAN, ProASN, TITAN, obj-sys, rasn, ffasn1 \\ \hline 
        S8 & Missing \textsf{SIZE} Extension Root Constraint Check$^\dagger$  & 20.4 & Strong-Inj & NC & asn1c, pycrate, asn1tools, rasn \\ \hline  
        S9 & Missing \textsf{INTEGER} Extension Root Constraint Check$^\dagger$ & 13.1 & Strong-Inj & NC & asn1c, pycrate, asn1tools, rasn \\ \hline 
        S10 & \textsf{ENUM} Extension Root Out-of-Bounds & 14.3 & Strong-Inj & NC & srsRAN  \\ \hline 
        S11 & \textsf{CHOICE} Extension Root Out-of-Bounds & 23.8 & Strong-Inj & NC & srsRAN  \\  \hline 
        S12 & Out-of-Bounds Bit-Cursor for Unknown Extension & 19 & Strong-Inj & SD & ProASN \\ 
        \hline    
    \end{tabular}
    \end{center}
{\small \textnormal{ \textbf{X.691}: the specification section entry that details this Non-Compliance. 
\textbf{Impact}: the potential effect on the overall system, which depends on the system’s assumptions about the parser. Vul: Vulnerability, NC: Non-Compliance, SD: Security Deviation; 
DoS: Denial of Service, MC: Memory Corruption,
\\ 
$\dagger$ These results exclude the ProASN and srsRAN parsers, as 5G/4G protocols lack these types, leaving their conformance unknown.  
}}

\end{table*}

By analyzing the discrepancies from testing results, we found 20 types of non-compliance (Table~\ref{tab:violation}), categorized by their respective correctness property violation (\cref{subsec:correctness_guarantees}).
We begin by discussing some common observations before detailing the specifics.

\smallpara{Non-compliances Among All Existing Parsers}
Our evaluations reveal 4 classes of non-compliances (S2, S5, S6, S7) that occur in all parsers tested. 
Notably, using standard differential fuzzing that relies on the consensus among existing UPER parsers would fail to identify these flaws. 
This highlights the inherent limitations of pure differential testing  
and the need for a trusted oracle to properly establish a ground truth.

\smallpara{Non-Compliance related to Extension}
A large portion of the non-compliance relates to handling extensions (e.g., R2, R4, S5, S6), as well as Open Type field. These findings empirically confirm that the extension mechanism represents a primary source of implementation complexity. This matches our observation in Challenge~\challengeref{1}. 

\smallpara{Oracle Limitations}
We acknowledge several limitations of our oracle. The class of false positives has already been discussed in \cref{subsec:vuper_accuracy}.
We identify three categories of false negatives. 
The first arises from missing features. If \system{} does not model a particular ASN.1 construct, it cannot flag violations involving that construct. 
The second category concerns mis-modeled features, where \system{} omits or incorrectly implements a particular check, especially when such checks are missed by both the oracle and all existing parsers. 
 While such cases are possible, we consider them unlikely, as the round-trip property limits the space for silently mis-modeled behavior.  
Finally, the fuzzer may fail to exercise certain edge cases in the input space. However, since the fuzzer itself is not our main contribution, we do not make claims about its completeness.

\smallpara{Security Impact} We categorize the security impact into three types: (a) \textit{Vulnerabilities:} Exploitable flaws (W1, W2, R4) as we demonstrated in \cref{subsec:attacks};  
(b) \textit{Security Deviations:} Flaws (R2-3, R5–6, S4, S12) that facilitate auxiliary threats, such as protocol confusion or device fingerprinting; and 
(c) \textit{Non-Compliance:} Violations (R1, S1–3, S5–11) that undermine canonicality and may risk interoperability failures.  

\subsubsection{Failure to comply with Weak-Injection}

\para{\hyperlink{W1}{W1}: Integer Constraint Check}
UPER encodes ranged integers in the minimum number of bits, which can leave more bit-states than range values: INTEGER (1..6) requires 3 bits, but 3 bits represent 8 values.
Our testing reveals that while some implementations, such as rasn~\cite{rasn}, omit the integer range check entirely, asn1c~\cite{ASN1c} generally enforces these checks but fails in the case of integer ranges exceeding $32$ bits with a positive lower bound (CVE-2025-55398). In this case, asn1c will compile the \textsf{INTEGER} type into an \texttt{unsigned long}, but the decoder fails to check the upper bound.

\para{\hyperlink{W2}{W2}: \textsf{SIZE} Constraint Check} Similarly, some implementations fail to enforce length constraints for \textsf{SEQUENCE OF}, \textsf{BIT STRING}, \textsf{OCTET STRING}, or other restricted character strings. 

\impact 
Omitting \textsf{INTEGER} range checks can trigger assertion failures when downstream applications validate the value, leading to DoS, or out-of-bounds array indexing, resulting in data corruption. Similarly, \textsf{SIZE} constraint violations can cause critical memory corruption, particularly when implementations rely on these constraints for memory allocation without validating dynamic input lengths. Such vulnerabilities are historically significant; for instance, a 4G RRC decoder in a Samsung baseband~\cite{hernandez_firmwire_2022} accepted invalid lengths, resulting in a stack-buffer overflow.

We find an illustrative legacy example in srsRAN 4G (v18.09). The \textsf{RRCReconfiguration} message contains a \textsf{drb-ToAddModList} field, which is defined to have a maximum of 11 items. The parser processes this by decoding a 4-bit length field into $n$ and iteratively populating a pre-allocated array of size 11. 
However, the implementation fails to validate whether $n$ falls in this range, allowing an attacker to supply a valid 4-bit integer exceeding 11 and trigger an immediate out-of-bounds memory write.

\subsubsection{Failure to comply with Surjection}

We identify two types of Surjection rule violations. 
First, the decoder may fail to decode a valid message.
For example, asn1c caps \textsf{SEQUENCE OF} types at 200 elements. 
Code inspection reveals this limit protects against SEQUENCE OF NULL compression bombs. 
This could cause interoperability issues, as valid messages might be rejected. 
Second, decoders may misinterpret inputs, i.e., accepting a message but mapping it to an incorrect internal representation.

\para{\hyperlink{R4}{R4}: Incorrect Handling of Unknown Extension} 
The srsRAN decoder fails to process unknown extensions correctly when an ASN.1 definition includes an extension marker (...) but no defined extension additions. When encountering a message from a future protocol version that contains extension data, the srsRAN decoder ignores the extension bit entirely and attempts to parse the subsequent field at the bit position where the extension should have been. This causes the decoder to misinterpret the remaining bitstream.

\para{\hyperlink{R5}{R5}: Incorrect Handling of Extension Count}
ProASN exhibits a similar problem in its handling of extension counts. According to the standard, the number of extensions is encoded as a normally small length determinant: a leading bit 0 indicates a count of up to 64, encoded in the subsequent 6 bits, while a leading 1 implies that an explicit length determinant follows. 
The issue arises that ProASN ignores the prefix bit, essentially always interpreting the next 6 bits as the count. Thus, if the sender includes more than 64 items in a message, ProASN would obtain an incorrect offset and misinterpret the remainder of the message.

\impact
Such deviations can cause interoperability failures between independently developed systems. 
A single field error can corrupt all subsequent decoding, creating logical inconsistencies for security-critical parameters. 
These failures can also lead to DoS if systems fail to establish or maintain a valid connection.

\subsubsection{Failure to comply with Strong-Injection}
These findings detail violations of the \textit{Strong-Injection} property caused by non-compliant ASN.1 implementations or lenient interpretations of the X.691 specification. Some violations could present interoperability risks and potential security vulnerabilities.

\para{Handling Difference} 
One issue is that vendors may handle messages inconsistently, with one system accepting a message that another rejects, resulting in potential unexpected behaviors. 
For example, the length determinant in Open Type field indicates how many bytes the encoding takes. However, when the length determinant mismatches the actual payload, for instance, a 3-byte declaration for 2 bytes of data, the parsers might differ in how they handle it. Pycrate~\cite{pycrate} would skip only 2 bytes in total and immediately start parsing the next field, while ProASN would skip 3 bytes in total and then start parsing the next field.

\para{Alternative Encodings}
Lenient parsers may accept multiple distinct encodings of the same value, violating non-malleability. 
For instance, unconstrained INTEGERs should not have the first 9 bits to be all zeros or all ones, as this indicates redundant leading octets.
Nonetheless, we find that many parsers accept positive values padded with extraneous leading \texttt{0x00} bytes.

\para{Non-Compliant Encoder}
\system{} also reveals non-conformant encoders. 
Beyond the encoding of the \textsf{DEFAULT} field by one of the network operators (\cref{subsec:vuper_accuracy}), we find similar issues in the srsRAN \cite{srsRAN_Project} and rasn \cite{rasn}. Per X.691 standard, an `ExtensionAdditionGroup' must be encoded as a missing extension addition if all its components are missing. 
However, these encoders fail to omit the group when it is marked present but contains only absent items. 

\impact 
First, \textit{strong-injection} violations could cause interoperability issues. When implementations exhibit differential handling of alternative encodings, a single broadcast message processed differently by receiving devices can lead to state desynchronization, mirroring the impact of surjection violations.
Similarly, a non-compliant encoder may emit messages that strictly conforming 
parsers reject, causing interoperability issues.
For instance, during 5G handover, if a source gNB incorrectly encodes a relayed Reconfiguration message from the target gNB, a UE employing a strict parser will discard the packet, resulting in handover failure. 

Second, these violations facilitate device fingerprinting. 
An active attacker with a fake base station can send specific alternative encodings to a device and observe which are accepted/rejected. The resulting accept/reject signature can often reveal the device manufacturer. 
For example, Flaw S12 is unique to ProASN in our findings. An attacker can craft an RRC message containing an unknown extension whose declared length exceeds the actual payload. If the device accepts and responds, the attacker can confirm that the target uses a Samsung baseband. 
Such fingerprinting further enables the attacker to exploit device-model specific known vulnerabilities, e.g., authentication bypass in Samsung Galaxy S21~\cite{tu2024logic}.

\begin{figure*}[t]
    \centering
    \begin{subfigure}{0.26\textwidth}
        \centering
        \includegraphics[width=\linewidth]{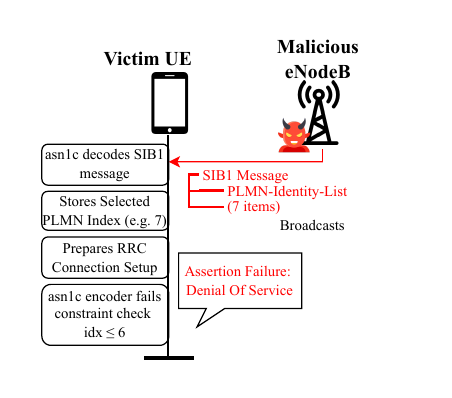}
        \caption{Out-of-Bound PLMN IdentityList causing UE crash.}
        \label{subfig:ue_crash}
    \end{subfigure}
    \hfill
    \begin{subfigure}{0.31\textwidth}
        \centering
        \includegraphics[width=\linewidth]{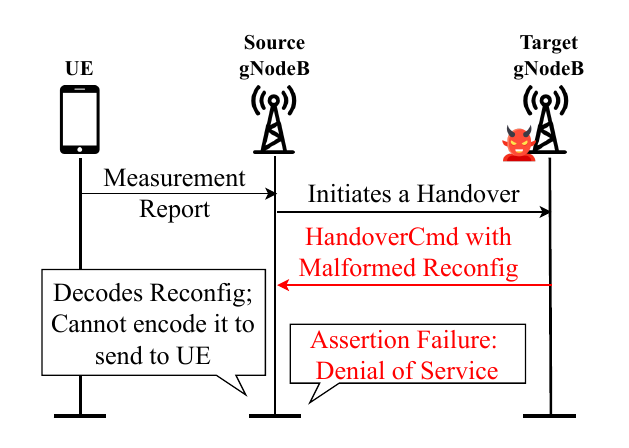}
        \caption{Constraint violation message causing gNB crash}
        \label{subfig:handover_gnb_crash}
    \end{subfigure}
    \hfill 
    \begin{subfigure}{0.4\textwidth}
        \centering
        \includegraphics[width=\linewidth]{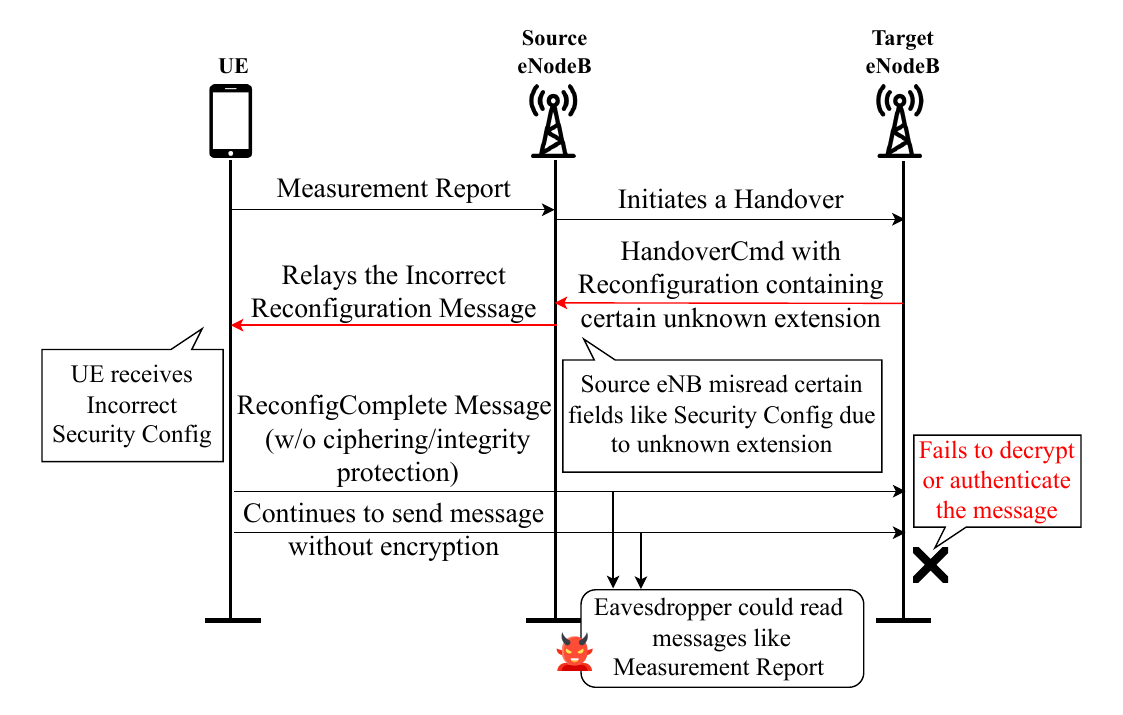}
        \caption{Miscommunication during handover}
        \label{subfig:handover_miscomm}
    \end{subfigure}
    \caption{Proof-of-Concept Attacks}
\end{figure*}

\subsection{Proof-of-Concept Attacks (Q3)} \label{subsec:attacks}
We present four proof-of-concept attacks demonstrating how low-level parser errors propagate to the downstream protocol layer and induce real-world security implications.

\subsubsection{Injection Violation Exploits} 
\label{subsec:injection_violation_exploits}
We identified two vulnerabilities in the OpenAirInterface \cite{OpenAirInterface} implementation of the 4G and 5G RRC layers by exploiting Flaw W1 and W2 of the asn1c.

\para{Out-of-Bound \textsf{PLMN IdentityList} causing UE crash}
The first issue, shown in Figure~\ref{subfig:ue_crash}, involves the PLMN (Public Land Mobile Network) Identity List. This is a critical data field within the SIB1 (System Information Block Type 1) broadcast from a cell tower to announce the carrier networks it supports. The 4G RRC specification limits this list to a maximum of 6.
However, a malicious eNB can broadcast a SIB1 message containing an oversized list with 7 or more identities, which the OAI UE stack fails to reject during initial parsing.
The UE then needs to store the index of its chosen carrier to maintain state across the asynchronous connection procedure.
This stored index is later used when the UE needs to inform the network which \textsf{PLMN} it has selected to complete the handshake. 
But the UE would fail to encode the stored index into the \textsf{selectedPLMN-Identity} field, defined as \textsf{INTEGER (1..6)}. 
This results in an assertion failure, allowing a malicious eNB to remotely crash the UE's RRC protocol stack.

\para{Constraint violation message causing gNB crash} The second vulnerability (Figure~\ref{subfig:handover_gnb_crash}) can be exploited during a 5G handover~\cite{5gRRC_NR} from the source $\text{gNB}_A$ to the target $\text{gNB}_B$. 
In this procedure, $\text{gNB}_A$ initiates a handover based on the UE Measurement Report and awaits a \textsf{HandoverCommand} from $\text{gNB}_B$. 
This command contains an \textsf{RRCReconfiguration} message intended for the UE to complete the handover. A compromised target node can inject a malformed \textsf{RRCReconfiguration} message that contains either a \textsf{SIZE} or \textsf{INTEGER} constraint violation like an overlong \textsf{DRB-ToAddModList}. 
Instead of rejecting the malformed packet during the initial decoding phase, $\text{gNB}_A$ accepts the packet and immediately attempts to re-encode it as a \textsf{DL-DCCH-Message} for over-the-air transmission. This re-encoding process fails as the encoder performs its own strict constraint checks; however, rather than gracefully handling the error, the system triggers a fatal assertion failure that crashes the program, resulting in a DoS on the source base station. 
A compromised femtocell \cite{femtocell} can cause such DoS by acting as the target gNodeB as shown in Figure \ref{subfig:handover_gnb_crash}.

\subsubsection{Surjection Violation Exploits} 
\label{subsec:surjection_violation_exploits}
Such surjection violations can lead to subtle yet critical miscommunications; although they may not trigger immediate system crashes, they can introduce severe logical errors. We discuss two such instances below. 

\para{Miscommunication between UE and gNB}
The first issue is a miscommunication during an \textsf{RRCReconfiguration} procedure between the UE and gNB within the srsRAN 5G implementation. Specifically, the \textsf{RadioBearerConfig} Information Element configures the keys for Data Radio Bearers (DRBs) using critical security fields, including \textsf{keyToUse} (selecting between the Master Key $K_{gNB}$ or Secondary Key $S-K_{gNB}$) and \textsf{sk-Counter} (providing freshness for secondary key derivation). 
We observe that certain message extensions could cause the UE to misinterpret critical security parameters, for example, setting \textsf{keyToUse} from the secondary key to the master key. This creates a mismatched security context and would cause the integrity and deciphering checks in DRBs to fail, effectively resulting in a session-terminating DoS.

\para{Miscommunication during eNB handover}
The second issue occurs during a 4G handover, as illustrated in Figure~\ref{subfig:handover_miscomm}. 
Similar to the 5G handover, the source eNB would signal the transfer by forwarding a \textsf{RRCReconfiguration} message from the target eNB to the UE. 
This message contains the specification of the new physical cell and critical \textsf{SecurityConfig} parameters, such as the selected encryption and integrity protection algorithms. 
We find that encodings involving unknown extensions can drive the UE into an inconsistent configuration state. As a result, the UE fails to synchronize with the target eNB due to invalid physical-layer parameters, causing handover failure. Moreover, inconsistencies in the interpreted security configuration can lead to a ciphering/integrity mismatch, prompting the target eNB to discard subsequent UE signaling. This bug may also induce information leakage if the UE mistakenly adopts a lower security level than intended by the network.

\subsection{Baseline Comparison}
To evaluate the unique contribution of \system{} as a verified test oracle, we compare its bug-detection capabilities against two baselines.
The first is \textit{negative testing}: deliberately generating invalid inputs for the parser to check whether it rejects them. The second is \textit{consensus testing}, i.e., differential testing without a verified oracle, where a bug is flagged when implementations disagree.

To enable a controlled comparison, we focus the baseline evaluation on 5 open-source implementations: asn1c, pycrate, rasn, asn1tools, and TITAN. Open-source parsers are well-suited for two reasons: they can be compiled against a single ASN.1 specification (whereas baseband and srsRAN parsers are tied to specific protocol versions), and source access makes it easier to triage. Because asn1tools and rasn do not support 5G formats, we evaluate on V2X protocols. This restricts the comparison to 14 of the 20 bug types found by \system{} (W1-2, R2-3, R6, S1-9), including the four bugs (S2, S5-7) present in all tested parsers.

\para{Negative Testing}
We use OTABase~\cite{park2025otabase}, an existing negative test case generator for the RRC layer. 
However, OTABase is originally designed to test the entire 4G RRC layer (i.e., both the parser and packet-processing logic) rather than strictly evaluating parsers. Hence, we adapt it and introduce the following changes.

First, we filter the mutation strategies to exclusively generate malformed inputs. We discard strategies that produce syntactically valid ASN.1 messages, like generating empty sequences that can test system boundaries, but fail to violate any parser constraints.
Concretely, this yields the following negative test case generation  strategies: 
(1) Out-of-range values for \textsf{INT/String};
(2) Out-of-set indices for \textsf{ENUM/CHOICE};
(3) Length-content mismatches in \textsf{SEQ-OF}; 
(4) Random bit truncation from valid packets.
Second, we extend OTABase to support V2X message generation.

Our evaluation generates over 10k test cases that cover all 4 types of generation strategies. 
Our testing reveals that OTABase can identify only W1 and W2, but misses the remaining vulnerabilities. 
This limitation arises primarily because OTABase focuses on testing RRC layer and mutating primitive data types and lacks the specific, schema-aware mutations necessary to trigger deep parsing bugs. For instance, ASN.1 extensions remain unexplored.  

\para{Consensus Testing}
We conducted a differential fuzzing campaign against the 5 aforementioned open-source parsers. 
By cross-referencing the accept/reject status and the decoded values for every input, we utilize the resulting discrepancies to identify non-compliance bugs. We, however, notice that a limitation of consensus testing is the triaging overhead, as we cannot use the semantic error codes from \system{}. 
We solve this challenge by tracking the semantic error code emitted by some parsers. However, they often force us to revert to the labor-intensive manual triaging.

Ultimately, this campaign identified five distinct bugs: W1, W2, R6, S8, and S9. 
This result is due to several reasons. First, differential testing inherently relies on implementation divergence, so non-compliances shared by all parsers are invisible. 
Second, this method suffers from parser-choice bias. It cannot expose bugs unique to untested parsers. 
Finally, this approach is subject to outlier shadowing. When a particular parser like rasn introduces a disproportionate number of behavioral differences, it adds noise to the triaging process and could mask other issues. 

Summarized in Table~\ref{tab:baseline_comparison}, nine bugs, including the four unique bugs (S2, S5, S6, and S7),  evade both negative and consensus testing, making them uniquely discoverable by \system{}.

\begin{table}[t]
    \caption{Bug Finding Baseline Comparison on V2X Protocols}
    \label{tab:baseline_comparison}
\scriptsize
\centering
\begin{tabular}{ | l  | l | l | }
\hline 
\textbf{Methodology} & \textbf{Logic Coverage} & \textbf{Found Bugs (V2X Subset)} \\ \hline 
Negative Testing & Primitive Bounds  & W1, W2 \\ 
Consensus Testing & Implementation Differences  & W1, W2, R6, S8, S9 \\
\textbf{\system{} as Oracle} & \textbf{Roundtrip-Property} &  \textbf{All 14 Types} \\
\hline 
\end{tabular}
\end{table}

\section{Discussion}

\para{Incorrect Usage of ASN.1 libraries} 
Our analysis also uncovered crashes in the RRC implementation of OAI~\cite{OpenAirInterface} resulting from unsafe assumptions in the application code. These issues are not inherent to the parser but are related to mixing protocol-level assumptions with the parser's ability to sanitize inputs.  
One (CVE-2025-55396) is an out-of-bounds access due to a mismatch in the size of a \textsf{SEQ-OF} between the 5G NR specification and the OAI's implementation. 
The other (CVE-2025-55397) involves a null pointer dereference where OAI assumes an optional RRC protocol field is always present.

\para{Bugs in ASN.1 Compiler} While our primary objective is not to audit ASN.1 compilers, our evaluation of diverse ASN.1 definitions revealed 2 compiler-level bugs. 
One bug is in rasn \cite{rasn} where the compiler erroneously generates $\text{import}$ statements for parameterized types that have already been unfolded. 
The second bug is in $\text{pycrate}$~\cite{pycrate} where the compiler crashes when a $\textsf{SEQUENCE-OF}$ type is defined with multiple constraints stacked together.

\para{Applicability to other Encoding Rules} 
Our methodology extends to other encoding rules such as APER, though adaptation poses challenges.
Because APER is not consistently byte-aligned, its encoding varies based on the current bit position, making the \textit{encode-invariance} property difficult to satisfy. Therefore, it is practical to focus exclusively on the weak-injection property.
Despite these differences, our findings directly transfer to APER: the asn1c~\cite{ASN1c} APER decoder also lacks \textsf{INTEGER} constraint checks. 

\para{Responsible disclosure}
We reported our findings to the respective library maintainers, receiving varied responses and 3 assigned CVEs, with several pending.
At least 3 bugs have already been patched, including asn1c's flaw \hyperlink{W1}{W1}, \hyperlink{W2}{W2} and pycrate's \hyperlink{R3}{R3}, with more detailed report in \extref{sec:disclosure}.
A notable discussion emerged over non-canonical compliance, where asn1c developers argue for the robustness principle~\cite{robustness}; other developers, e.g., Objective System welcomes our detailed edge-case examples, viewing them as valuable test cases for their implementation of the Canonical UPER decoder.
This reinforces \system{}'s potential as a reference implementation for improving parser security.

\section{Related Work}

\para{Verified Parser} 
Verified parsers are an active area of research that bridges formal verification, programming languages, and security. 
EverParse~\cite{Everparse19,swamy2022hardening} and Comparse~\cite{wallez2023comparse} are non-malleable parser libraries built in F*~\cite{fstar}. EverParse generates efficient, zero-copy C code, while Comparse focuses on integration with cryptographic frameworks like DY*~\cite{bhargavan2021dy}.
Rocq~\cite{Coq-refman} is used in both general frameworks like Narcissus~\cite{Narcissus19} and specific applications like the Protocol Buffer compiler~\cite{ye19verified}.
Other notable works include
Vest~\cite{vest_25}, a verified Rust parser using Verus~\cite{verus23}; 
PulseParse~\cite{pulse_parse}, a non-malleable parser for CBOR formats;
and Bigrammar~\cite{tan2018bidirectional}, an instruction decoder that satisfies \textit{surjection} and \textit{weak-injection}.

\para{ASN.1 Parsers}
The widespread use of ASN.1 has prompted many formalization efforts, primarily targeting the Distinguished Encoding Rules (DER), a byte-aligned encoding fundamentally different from UPER. ASN1$^\star$~\cite{ni2023asn1} uses EverParse~\cite{Everparse19} to generate non-malleable DER parsers.
Formalization of X.509~\cite{ITU_X509} has also yielded verified DER parsers as a byproduct~\cite{debnath2024armor, verdict_x509_25}.
Beyond DER, existing efforts include a UPER parser for vehicle communications~\cite{tullsen2018cav}, but it lacks essential features, and a verified parser for the non-standard, legacy ASN.1/ACN scheme~\cite{asn1acn25bucev} with no injection property. 
A comprehensive comparison of these projects is in Table~\ref{tab:verified_parser_comparsion}.

\para{Differential Testing \& Fuzzing}
Differential testing~\cite{mckeeman1998differential} has been successfully applied for cryptographic libraries~\cite{cryptofuzz}, 
X.509 validation~\cite{ceres21, verdict_x509_25, petsios2017nezha, debnath2024armor} and ZIP parser \cite{zip_diff}. 
Our approach of using a trusted reference is similar to that of ARMOR~\cite{debnath2024armor} or Verdict~\cite{verdict_x509_25}.

\section{Conclusion}
In this paper, we present \system{}, a verified parser/serializer framework for ASN.1 UPER, one of the widely used interface description languages for complex networks. \system{} provides bit-level reasoning and round-trip properties to ensure the parser correctness and security, support for forward/backward compatibility, and verified ASN.1 combinators. 
We develop a testing framework using \system{} as a verified oracle. Our experiments uncovered 20 types of inconsistencies in 7 open source and 4 commercial implementations. This work highlights the need for more robust, high-assurance parsers.

\begin{acks}
We thank the anonymous reviewers and the shepherd for their feedback and suggestions. We also thank the corresponding developers for cooperating with us during our responsible disclosure. This paper was edited for grammar using Claude and Gemini. This work has been supported by
the NSF under grants 2145631, 2215017, and 2614551, and the Public Wireless Supply Chain Innovation Fund (PWSCIF) under Federal Award ID Number 51-60-IF007.
\end{acks}

\bibliographystyle{ACM-Reference-Format}
\bibliography{bibliography/software, bibliography/ref, bibliography/techreport}

\appendix

\section*{Ethical Considerations}

This work develops a formally verified ASN.1 UPER parser and uses it to uncover non-conformance in major implementations. The principal ethical risk is that disclosing such flaws could aid attackers. We mitigated by reporting every finding to the respective maintainers with proof-of-concept inputs and offering assistance with remediation. The real-world impact (e.g., DoS, authentication bypass) depends on how each parser is integrated downstream, and we provide no weaponized exploits, limiting risk to end users. All experiments were legal and run in isolated, self-hosted testbeds, affecting no external systems or live networks. We assess that the security benefits to parser developers, dependent vendors, end users, and the research community outweigh these mitigated risks.

\section*{Open Science}

To support open science, we release our artifacts publicly.\footnote{\url{https://github.com/SyNSec-den/VUPER}} The release includes \system{}'s verified parser/serializer, the parser compiler and the differential-testing harnesses.

\ifarxiv

\section{Additional Findings}
\label{sec:additional_findings}
Here, we discuss some additional findings. 

\para{Additional Surjection Violation} 

\smallpara{\hyperlink{R3}{R3}: Incorrect Handling for UTF8String Constraint}
It is a case where constraints on UTF8String types are not \textit{PER-visible} and should therefore be ignored during encoding and decoding. However, implementations such as pycrate incorrectly enforce these rules. Specifically, instead of correctly decoding a length determinant, pycrate mistakenly decodes a \textsf{SIZE} constraint first, which is represented similarly to a constraint \textsf{INTEGER}, creating a deviation from the standard. A concrete example is given in Figure~\ref{fig:utf8_encode}. 

\smallpara{R6: Incorrect Handling of Normally Small Length Det} 
In rasn~\cite{rasn}, the handling of a normally small length determinant is incorrect. It handles the determinant as if it is a normally small integer. This would cause their decoding to fail or diverge from the actual message. 

\begin{figure}[h]
    \centering
    \includegraphics[width=\linewidth]{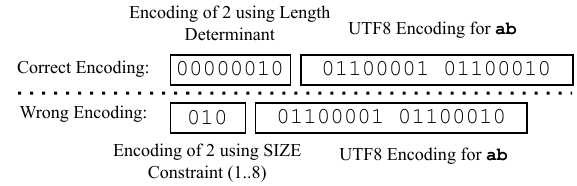}
    \caption{Wrongful Encoding from Pycrate when encoding string \texttt{ab} for \texttt{UTF8STRING (0..7)}}
    \label{fig:utf8_encode}
\end{figure}

\para{Additional Strong-Injection Violation}
\begin{description}[font=\itshape]
    \item[Mismatched Extension Root Constraint Checks.] For example, \textsf{INTEGER (1..10, ...)}. There are two levels of this problem. First, if a value is encoded as a non-extension, it should be between 1 and 10; some ASN.1 libraries will not enforce it. Second, if a value is encoded as an extension, it should not be between 1 and 10.  

    \item[Extension Addition Group.] The standard requires that ``If all components values of the ``ExtensionAdditionGroup'' are missing then, the ``ExtensionAdditionGroup'' shall be encoded as a missing extension addition''. Therefore, if a message encodes the Extension Addition Group, but all the fields are empty, it should be considered wrongly encoded. 

    \item[Encoding of Default Value.] Default value is unnecessary when a default value is provided to a default field, and it is a simple type, according to X.691. Consequently, if a default field is encoded but turns out to have the default value, the encoding should be rejected.

    \item[Wrongful encoding of Choice/Enum extension.] Here, the issue is that the input can overflow the tag of the choice, then the parser wrongly assumes it will be extensions, and decode them. However, this contradicts the ASN.1 logic. 

    \item[Overlong length determinants.] In the encoding rule, any length less than 128 should occupy one octet, and a length between 128 and 16384 should occupy 2 octets. Figure \ref{fig:over-long} shows an example of an overlong encoding for the length determinant. A similar case exists for the normally small length determinant.

\end{description}

\begin{figure}[h]
    \centering
    \includegraphics[width=\linewidth]{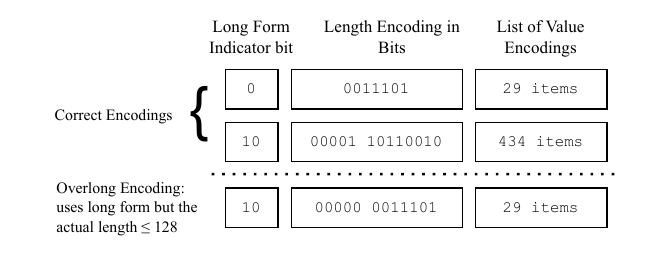}
    \caption{Length Determinant Overlong Encoding}
    \label{fig:over-long}
    {Since the length of 29 is less than 128, the rules mandate using the single-byte short form. An overlong encoding improperly uses the two-byte long form for this value, creating a non-canonical representation. }
\end{figure}

\subsection{Performance Evaluation}
\label{subsec:performance}
To evaluate \system{}'s performance against other implementations, we run the compiled decoder on an Ubuntu 24.04 machine equipped with an Intel Core i9-14900K and 32 GB of memory.
We run the decoder on a synthetic dataset of 6394 \textsf{DL-DCCH-Message}s and 9501 \textsf{UL-DCCH-Message}s 100 times and collect the average execution time. Our parser is on average 4x slower than C implementations and 4x faster than pycrate (Table \ref{tab:performance}). 
The performance overhead of \system{} is primarily due to OCaml’s garbage collector runtime and the use of certain high-level data structures, such as \textsf{list bool} for \textsf{SEQUENCE} bitmaps, which prioritizes verification simplicity over execution speed. Optimizing these structures or extracting code directly to C~\cite{erbsen2020fiat} remains future work.

\begin{table}[h]
    \caption{Performance Evaluation}
    \label{tab:performance}
    \centering
    \scriptsize
    \begin{tabular}{ c | c | c | c | c | c }
        \toprule \hline 
        \textbf{ASN1 tool} & ASN1c & Pycrate & TITAN & Obj-Sys & \textbf{\system{} Parser} \\
        \hline 
        \textbf{Language} & C & Python & C++ & C & OCaml \\ \hline 
        \textbf{Time} \textsf{DL} (s) & 0.0365 & 0.6281 & 0.0353 & 0.0365 & \textbf{0.1731}  \\ \hline 
        \textbf{Time} \textsf{UL} (s) & 0.0645 & 1.1234 & 0.0853 & 0.0600 & \textbf{0.2137} \\  
        \hline \bottomrule
    \end{tabular}
\end{table}

\section{Additional Definitions for Parser Properties}
\label{sec:additional_definitions}
In this section, we will give a primer on Rocq~\cite{Coq-refman}, and list the definitions of some of the properties mentioned in \cref{subsec:correctness_guarantees}.  We will  add some examples for \cref{subsec:parser_combinators}.

\subsection{Primer on Rocq}
Rocq, formerly known as the Coq Theorem Prover, is based on the Calculus of Inductive Constructions, a type theory that views Propositions as Types. 
This means that a mathematical proposition is represented as a type, and a proof is simply a program that has that type.

Rocq represents multi-argument functions in curried form. For example, the function $f : \mathbb N \times  \mathbb N \times  \mathbb N \to \mathbb N $ is viewed as a sequence of functions that each take a single argument. Namely, 
$f :\mathbb N \to  \mathbb N \to  \mathbb N \to  \mathbb N $. 

We utilize Sigma Types like $\{x : A  | P x\}$. These are dependent pairs, which consist of a value $x$ of type $A$ and a proof (evidence) that $x$ satisfies the logical condition $P$. In the example, \textsf{INTEGER(1..8)} can correspond to 
\begin{lstlisting}[style=mathstyle]
Definition range (x : $\mathbb Z$) : Prop := 1 $\leq$ x $\leq$ 8.
Definition constrained_int : Type := {x : $\mathbb Z$ | range x}.
Lemma pf_2 : 1 $\leq$ 2 $\leq$ 8.
Proof. lia. Qed. 
Definition v : constrained_int := exist range 2 pf_2. 
\end{lstlisting}
Here, \textsf{v} contains the number 2 and the lemma \textsf{pf\_2}. And \textsf{lia} is a powerful Rocq tactic that solves Linear Integer Arithmetic.

We assume proof irrelevance, meaning that for any proposition $P$, any two proofs $\textsf{pf}_1, \textsf{pf}_2 : P$ are considered identical. This is an assumption on the proofs and does not affect the semantic outcome of the parser.

\subsection{Additional Property Definitions}

\para{Decode Invariance} It states that if decoding \( b_1 \) yields \( a \), then decoding a bit-equivalent buffer \( b_2 \) should produce the same result. This will ensure that the parser is not dependent on unrelated factors like position.
\begin{lstlisting}[style=mathstyle]
Definition decode_invariance :=
$\forall$ ($b_1$ : $\B$) ($p_1$ $p_1'$ : $\Pos$) ($a$ : $A$) (pf : $P$ $a$) (flg : Flg), 
  dec $b_1$ $p_1$ = Some (exist $P$ $a$ pf, $p_1'$, flg) $\to$
  ($\forall$ ($b_2$ : $\B$) ($p_2$ $p_2'$ : $\Pos$), equiv_bits $b_1$ $b_2$ $p_1$ $p_1'$ $p_2$ $p_2'$ $\to$
     dec $b_2$ $p_2$ = Some (exist $P$ $a$ pf, $p_2'$, flg))    $\land$ ($p_1$ $\leq$ $p_1'$).
\end{lstlisting} 

\para{Length-Consistency} This ensures the length function \textsf{msg\_len} accurately computes the encoding length as the difference between final and initial positions.

\begin{lstlisting}[style=mathstyle]
Definition length_consistency := 
  $\forall$ ($a$ : $A$) ($b$ $b'$ : $\B$) ($p$ $p'$ : $\Pos$) (pf : $P$ $a$),
    enc $b$ $p$ $a$ = Some ($b'$, $p'$, pf) $\to$
    msg_len $a$ = Some (to_nat $p'$ - to_nat $p$).
\end{lstlisting} 

\para{Format Correct} This property is defined to be the conjunction of all correctness rules, which can be formally defined as 
\begin{lstlisting}[style=mathstyle]
Definition format_correct 
  (enc : Serialize $A$ $P$)
  (dec : Parse $A$ $P$)
  (msg_len : $A$ $\to$ option $\mathbb N$) := local_write enc $\land$ encode_invariance enc $\land$ 
    decode_invariance dec $\land$ surjection enc dec $\land$ 
    strong_injection enc dec $\land$ weak_injection enc dec $\land$ 
    length_consistency enc msg_len. 
\end{lstlisting}
And Fmt is defined as follows, 
\begin{lstlisting}[style=mathstyle]
Record Fmt ($A$: Set) ($P$ : $A$ $\to$ Prop) := {
  enc : Serialize $A$ $P$;  dec : Parse $A$ $P$;
  msg_len : $A$ $\to$ option $\mathbb N$ ;
  correct : format_correct enc dec msg_len }.
\end{lstlisting}

\subsection{Additional Example on Parser Combinator}
\para{Basic $n$ Bit}
This is the type for the put/read $n$ bit functions. 
We use \textsf{nat} since Rocq's primitive types are difficult to reason about. 
The parameter $n$ represents the bit length. The decoder and encoder type are defined as follows:
\begin{lstlisting}[style=mathstyle]
put_n_bits ($n$ : nat) : Serialize nat ($\lambda$ $x$ $\Rightarrow$ $x < 2^n$)
read_n_bits ($n$ : nat) : Parse nat ($\lambda$ $x$ $\Rightarrow$ $x < 2^n$)
\end{lstlisting}

\para{Bool Format }
This is the definition of boolean format, which we defined using a combination of \textit{basic\_bit} and \textit{map}. 
\begin{lstlisting}[style=mathstyle]
Definition f1 ($x$ : bool) := if $x$ then 1 else 0. 
Definition f2 ($y : \mathbb N$) := if $y$ $=?$ 0 then false else true.
Lemma pf : $\forall$ $x$ : bool, f2 (f1 $x$) = $x$ $\land$ $\forall$ $y : \mathbb N$, $y < 2$ $\to$ f1 (f2 $y$) = $y$.
(* Proof is omitted *)
Definition bool_format : Fmt bool ($\lambda$ _ $\Rightarrow$ True) 
  := map (basic_bit 1) f1 f2 pf. 
\end{lstlisting}
Supplying a formal inversion proof $\textsf{pf}$ enables the \textit{map} to derive the new format.

\subsection{Compiler Workflow} 
\label{subsec:compiler_workflow}
Our compiler processes an ASN.1 definition in two main phases, shown in Figure~\ref{fig:compiler_generated_type}. 
First, the compiler recursively generates the Rocq data types, starting with component fields and ending with the final structured \textsf{Record} type and its refinement conditions.
Second, it generates the verified formats, recursively composing individual field formats with the ASN.1 combinator (like $\textsf{seq\_ext}$). This process yields an internal data structure organized as a nested product of the field types, rather than the final \textsf{Record} type. Next, the compiler adds a final \textit{map} format that maps the internal tuple representation onto the user-facing \textsf{Record} type.

\begin{figure}
    \centering
    \begin{lstlisting}[style=mathstyle,basicstyle=\footnotesize\sffamily]
Inductive xOverhead_Type := | xOh6 | xOh12 | xOh18.
(* Omit other intermediate types like ServCellIndex *)
Record PDSCH_ServingCellConfig_Type : Set := {
  codeBlockGrpTx : 
    option (SetupRelease_Type PDSCH_CodeBlockGrpTx_Type);
  xOverhead : option xOverhead__Type ;
  nrofHARQ : option nrofHARQ_Type ;
  pucch_Cell : option ServCellIndex_Type ;
... (* Extensions *) }.
(* A group of intermediate definitions for helper formats *)
Definition PDSCH_ServingCellConfig_Format : 
  Fmt PDSCH_ServingCellConfig_Type     PDSCH_ServingCellConfig_Condition := (* ... *)
\end{lstlisting}
    \caption{The Rocq type for \textsf{PDSCH-ServingCellConfig}}
    \label{fig:compiler_generated_type}
\end{figure}

\section{Additional Background on ASN.1}
\label{sec:additional_background}

\para{Details on Basic UPER and Canonical UPER}
In practice, most protocols opt to use the less restrictive Basic UPER, which is essentially bijective and closely aligns with Canonical UPER, with a few exceptions. 
One notable exception is \textsf{SET-OF}, which requires elements to be sorted to ensure a unique encoding. Because of this added complexity, it is almost never used in practice; for example, most network protocols, including 4G/5G~\cite{LTE_RRC, 5gRRC_NR} and V2X~\cite{CAM_ITS, DENM_ITS}, avoid it and instead use \textsf{SEQUENCE-OF} as the simpler alternative.
Hence, our work focuses on common real-world protocol constructs and aims for a clean formalization. We thus exclude \textsf{SET-OF} from our formal model and similar corner cases from our UPER model. This ensures a natural and precise application of the bijective properties.

\para{\textsf{SetupRelease} Definition}
The definition of a SetupRelease is as follows: it is essentially an option type. 
\begin{lstlisting}[style=asn1style]
SetupRelease { ElementTypeParam } ::= CHOICE {
 release NULL, setup ElementTypeParam }
\end{lstlisting}

\para{JER Output}
Figure \ref{subfig:example_msg_a}'s message encodes in JER as follows: 
\begin{lstlisting}[style=asn1style]
{
  "nrofHARQ": "n16",
  "maxMIMO-Layers": 2
}
\end{lstlisting}

Certain parsers have different ways to represent extension fields. 
For example, asn1c~\cite{ASN1c} would print the example message as 
\begin{lstlisting}[style=asn1style]
{
  "nrofHARQ": "n16",
  "ext1" : {  "maxMIMO-Layers": 2  }
}
\end{lstlisting}
This requires an extra step of normalizing the JER output.

\para{Baseband Parser Testing}
As for the baseband parsers, we use reverse engineering to identify the underlying meaning of each field (both inputs and outputs) to pinpoint the return value or field that represents the number of bits read. 
While there is a field that stores the decoded value, its type is unrecoverable.

\section{Additional Experiment Data on Forward/Backward Compatibility}
\label{sec:additional_compatibility_experiment}
\para{Examples of Structural Redefinition} 
We will show one example of a structural redefinition. 
In Release 16, \textsf{RRCReestablishment-IEs} is defined as 
\begin{lstlisting}[style=asn1style]
RRCReestablishment-IEs ::=          SEQUENCE {
    nextHopChainingCount                NextHopChainingCount,
    lateNonCriticalExtension            OCTET STRING                    OPTIONAL,
    nonCriticalExtension                SEQUENCE {}                     OPTIONAL
}
\end{lstlisting}
But in Release 17, it is redefined as, 
\begin{lstlisting}[style=asn1style]
RRCReestablishment-IEs ::=          SEQUENCE {
    nextHopChainingCount                NextHopChainingCount,
    lateNonCriticalExtension            OCTET STRING                    OPTIONAL,
    nonCriticalExtension                RRCReestablishment-v1700-IEs    OPTIONAL
}
RRCReestablishment-v1700-IEs ::=    SEQUENCE {
    sl-L2RemoteUE-Config-r17    SetupRelease {SL-L2RemoteUE-Config-r17} OPTIONAL,
    nonCriticalExtension        SEQUENCE {}                             OPTIONAL
}
\end{lstlisting}
The remaining 6 types that are changed are: 
\begin{sloppypar}
\noindent \textsf{LoggedMeasurementConfiguration-r16-IEs, DLInformationTransfer-v1610-IEs, RRCReconfiguration-v1610-IEs, RRCRelease-v1610-IEs, RRCResume-v1610-IEs, UEInformationRequest-r16-IEs}. 
\end{sloppypar}

\para{Examples of Extension Messages} 
We give a part of a \textsf{DL-DCCH-message} that is accepted by both Rel-16 and Rel-17 parsers. Rel-17 outputs the following: 
\begin{lstlisting}[style=asn1style]
"measObjectNR": {
    "associatedMeasGapCSIRS-r17": 8,
    "offsetMO": {
        "rsrpOffsetCSI-RS": "dB0",
        -- omitted
    },
    "quantityConfigIndex": 2,
    "referenceSignalConfig": {}
}
\end{lstlisting}
while the Rel-16 parsers give the result while flagging a \textsf{DiffVer}: 
\begin{lstlisting}[style=asn1style]
"measObjectNR": {
    "offsetMO": {
        "rsrpOffsetCSI-RS": "dB0",
        -- omitted
    },
    "quantityConfigIndex": 2,
    "referenceSignalConfig": {}
}
\end{lstlisting}
This is expected since \textsf{associatedMeasGapCSIRS-r17} is an extension field from Release 17.


\section{Responsible Disclosure Details}
\label{sec:disclosure}
We reported every finding to the corresponding maintainer or vendor with proof-of-concept inputs and
offered assistance with remediation.
asn1c confirmed and patched both weak-injection flaws, the missing \textsf{INTEGER} range check
(\hyperlink{W1}{W1}, CVE-2025-55398) and the missing \textsf{SIZE} constraint check
(\hyperlink{W2}{W2}); for the strong-injection findings its maintainers initially invoked the
robustness principle~\cite{robustness}, and the outcome was a canonical decoding flag that rejects
the non-canonical encodings we reported while leaving the default behavior unchanged.
pycrate fixed its \textsf{UTF8String} constraint error (\hyperlink{R3}{R3}), and rasn fixed its
handling of the normally small length determinant (\hyperlink{R6}{R6}).
Samsung acknowledged four findings in the baseband parser we denote ProASN: the extension count
(\hyperlink{R5}{R5}), Open Type length validation (\hyperlink{S4}{S4}), explicitly encoded
\textsf{DEFAULT} values (\hyperlink{S7}{S7}), and the out-of-bounds bit cursor on unknown extensions
(\hyperlink{S12}{S12}).
Objective Systems and ffasn1 welcomed our edge-case inputs as test cases for their own decoders.

\fi

\end{document}